\documentstyle[11pt,Moriond,epsfig,wrapfig]{article}

\newcommand{\ee}{\mbox{$\mathrm{e}^{+}\mathrm{e}^{-}$}}

\newcommand{\pb}{\mbox{$\mathrm{pb}^{-1}$}}

\newcommand{\mZ}{\mbox{$m_{\mathrm{Z}}$}}
\newcommand{\mH}{\mbox{$m_{\mathrm{H}}$}}

\newcommand{\mh}{\mbox{$m_{\mathrm{h}}$}}
\newcommand{\mA}{\mbox{$m_{\mathrm{A}}$}}
\newcommand{\mHpm}{\mbox{$m_{\mathrm{H}^{\pm}}$}}

\newcommand{\mW}{\mbox{$m_{\mathrm{W}^{\pm}}$}}

\newcommand{\Gcs}{\mbox{${\rm GeV}/c^2$}}

\def\be{\begin{equation}}
\def\ee{\end{equation}}
\def\bea{\begin{eqnarray}}
\def\eea{\end{eqnarray}}

\begin{document}
\unitlength 1cm
\vspace*{.5cm}
\title{Cornering Higgs Bosons at LEP~\footnote{Talk given at the XXXV$^{\scriptsize \hbox{th}}$ Rencontres de Moriond QCD and Hadronic interactions.}}

\author{Marumi Kado}

\address{European Laboratory for Particle Physics (CERN),\\
1211 Geneva 23, Switzerland.}

\maketitle\abstracts{
The most recent results of the searches for Higgs bosons in the 
data taken at LEP in 1999 and their interpretations in
a variety of theoretical frameworks are reviewed. }

In 1999 LEP reached a centre-of-mass energy of 202~GeV and delivered 
a total integrated luminosity  
in excess of 1~fb$^{-1}$ to the four experiments. A 
review of the results of all Higgs boson searches 
performed with the data collected in 1999 
is presented, with a particular emphasis on the combinations of the four 
LEP experiments. When for a given search such a combination is not available, 
the 
individual result with the best sensitivity is presented instead.

\section{The Standard Model}\label{subsec:msm}

In the standard model, the spontaneous breaking of the electroweak symmetry
is achieved at the expense of the introduction of multiplets
of complex scalar fields in self-interaction. In particular, in its 
minimal version, a single scalar doublet is required (with the two fold 
advantage of simplicity and of preserving $\rho_o=1$). As it develops 
a vacuum expectation value, gauge bosons acquire their masses absorbing
three of the four initial degrees of freedom and a massive scalar 
physical state remains, the Higgs boson h. Its mass is a free parameter
of the theory as a function of which its production and decays are predicted 
unambiguously.

%\begin{table}[t]
%\caption{Integrated luminosities for each individual experiment collected 
%in 1999 at.\label{tab:lumi}}
%\vspace{0.4cm}
%\begin{center}
%\begin{tabular}{|c|c|c|c|c|}
%\hline
%$\sqrt{s}$ GeV & ALEPH & DELPHI & L3 & OPAL \\ \hline
%192 & 28.9 & 25.2-25.9 & 29.7 & 28.7-28.9 \\ \hline
%196 & 79.9 & 74.8-76.9 & 83.7 & 73.9-74.8 \\ \hline
%200 & 86.3 & 81.9-84.3 & 82.8 & 74.8-77.2 \\ \hline \hline
%202 & 41.9 & 40.0-41.1 & 37.0 & 35.2-36.1 \\ \hline \hline
%Total &237.0 & 221.9-228.2 & 232.4 & 212.6-217.0 \\ \hline
%\end{tabular}
%\end{center}
%\end{table}

\begin{wrapfigure}{r}{6cm}
\vspace{-.2cm}
\mbox{\epsfxsize=1.\hsize\epsfbox{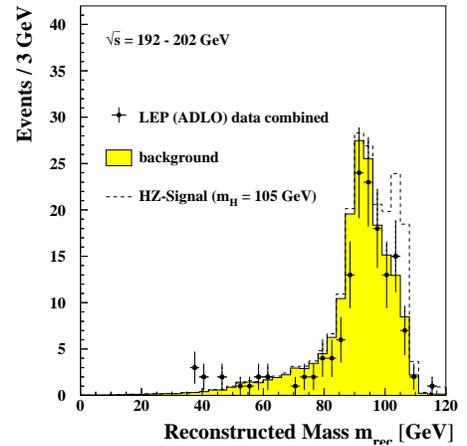}}
\caption{Distribution of the reconstructed mass of the Higgs boson 
at all energies above 192 GeV.
A restricted set of data among the selected candidates (dots) 
amounting to 148 events is used; 175 events are expected from
background processes 
(shaded histogram). The 36 events expected from 
a 105~\Gcs\ Higgs boson signal 
are also displayed (dashed histogram). 
\label{msm:mass}}
\vspace{-.8cm}
\end{wrapfigure}
At LEP, the minimal standard model Higgs boson dominant production process is
the Higgs-strahlung ($\mathrm{e}^{+}\mathrm{e}^{-} \rightarrow {\rm HZ}$) and 
its prominent decay is into a pair of b quarks ($\sim$ 81\% for a 
105~\Gcs\ Higgs boson). Four 
search topologies arise from the decays of the 
accompanying Z boson. 1) Four jets, when the Z decays into a pair 
of quarks; 2) two acoplanar jets and missing energy, when the Z  
decays to a pair of neutrinos; 3) the leptonic channel, when the
Z decays to a pair of electrons or muons; and 4) the $\tau\tau\hbox{q}\overline{\hbox{q}}$ and $\hbox{b}\overline{\hbox{b}}\tau\tau$ channels, where 
either the Z or the Higgs boson decay to a pair of taus. In all these 
channels but the latter, the Higgs boson is 
searched for in its $\hbox{b}\overline{\hbox{b}}$ decay mode. 
The b jet tagging is therefore an essential component of all
the analyses.

%\begin{wrapfigure}{l}{7.8cm}
%\vspace{-.3cm}
%\begin{tabular}{|c|c|c|c|c|}
%\hline
%Experiment & A & D & L & O \\ \hline \hline
%Candidates & 74 &282 & 59 & 53 \\ \hline 
%Background & 97.8 & 307.5 & 49.8 & 51.5 \\ \hline \hline
%Sensitivity & 107.7 & 106.3 & 105.3 & 105.2 \\ \hline
%Limit & 107.7 & 103.9 & 106.0 & 103.0  \\ \hline
%\end{tabular}
%\vspace{-.3cm}
%\end{wrapfigure}

All the data collected have been analysed by all four 
experiments~\cite{lephwg}, nevertheless the most significant 
contribution for the limit are the $\sim$40\pb/experiment collected 
at 202 GeV. The distribution of the reconstructed mass of the Higgs boson 
for all
channels and all experiments together is illustrated in 
Fig.~\ref{msm:mass}.
Only the most signal-like events are chosen with the further requirement 
that the contribution from all experiments be roughly the same.

\begin{figure}[h]
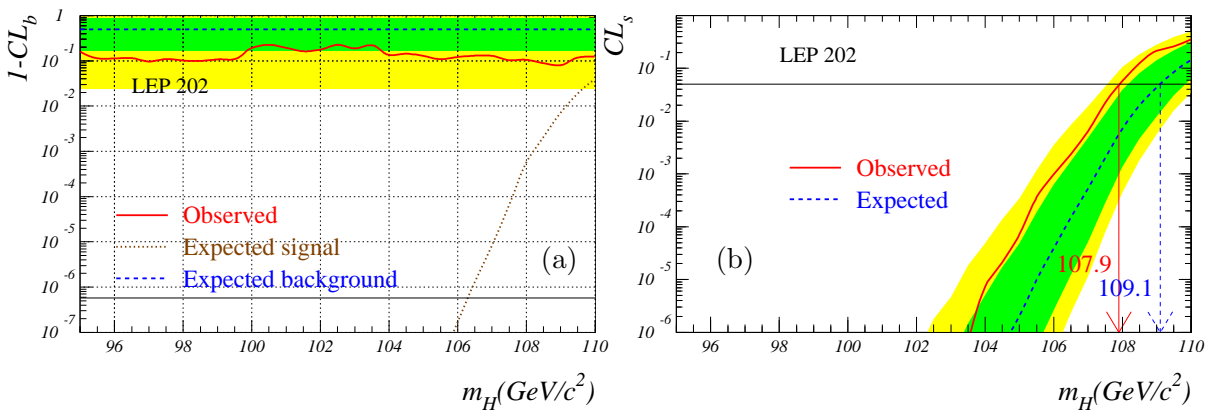

\begin{center}
\begin{tabular}{cc}
\mbox{\epsfxsize=.5\hsize\epsfbox{lep202_mclb.epsi}}  
\put(-.9,1.9){(a)}& 
\hspace{-.5cm}\mbox{\epsfxsize=.5\hsize\epsfbox{lep202_cls.epsi}}
\put(-6.5,1.9){(b)} 
\end{tabular}
\caption{The $CL_b$ (a) and $CL_s$ (b) confidence levels as a function of \mH.
The expected and observed values are indicated by dashed and plain lines 
respectively. The $\pm 1 \sigma$ and $\pm 2 \sigma$ probability bands off
the expectation in the absence of signal are illustrated as shaded regions. 
The dotted line in (a) represents the 
expected $CL_b$ in the presence of signal 
and the horizontal thin line at $5.7 \times 10^{-7}$ indicates the 5$\sigma$ 
discovery level.  \label{msm:cl}}
\end{center}
\end{figure}

The observed LEP combined $CL_b$ (which estimates the probability that the 
data be compatible with the background)~\cite{jadib} as a 
function of the Higgs boson mass 
hypothesis is illustrated in Fig.~\ref{msm:cl}a.
The expected values of $CL_b$ in the presence of signal are also shown in 
Fig.~\ref{msm:cl}a, yielding a 5$\sigma$ discovery sensitivity of 
106.3~\Gcs (corresponding to $CL_b \sim 5.7 \times 10^{-7}$).

In Fig.~\ref{msm:cl}b the $CL_s$ (which estimates the probability that the 
data be compatible with background and signal)~\cite{jadib}, computed 
without taking
into account the correlations among systematic errors, is illustrated as a 
function of the Higgs boson mass hypothesis. Hypotheses yielding a 
$CL_s < 5$\% are excluded at the 95\% confidence level. The sensitivity of the 
combination is 109.1~\Gcs. According to Fig.~\ref{msm:cl}b, Higgs boson mass
hypotheses below 107.9~\Gcs\ are excluded at the 95\% CL. However 
the effect of correlations of the systematic errors 
result in a downward shift of 150~MeV/c$^2$
of the limit, leading to combined lower limit on the mass of the Minimal 
Standard 
Model Higgs boson of 107.7~\Gcs.

\section{Extending the Higgs Sector of the Standard Model}

The simplest extension of
the minimal standard model Higgs sector is obtained with the introduction 
of an additional doublet of complex scalar fields 
(still preserving $\rho_o=1$). 
In two-higgs-doublet models (2HDM) FCNC can become important, but can 
be avoided if all
fermions of a given charge do not couple to more than one Higgs 
doublet. The 
two most popular 
models obeying this requirement are the model~I, where
one Higgs doublet couples only to fermions and the other to bosons, and 
the model~II where one Higgs doublet couples only to up-type fermions and the 
other to down-type fermions. In both cases, the electroweak symmetry 
breaking leaves five massive scalar physical states among which two are
neutral CP-even (h and  H), one is neutral CP-odd (A) and two are
charged  (H$^\pm$). These models are the main motivation for the search for
Higgs bosons decaying mostly to photons (model~I) and for charged Higgs bosons.

\subsection{2HDM of type I, fermiophobia \label{sec:fermio}}

\begin{figure}[h]
\begin{center}
\begin{tabular}{cc}
\mbox{\epsfxsize=.45\hsize\epsfbox{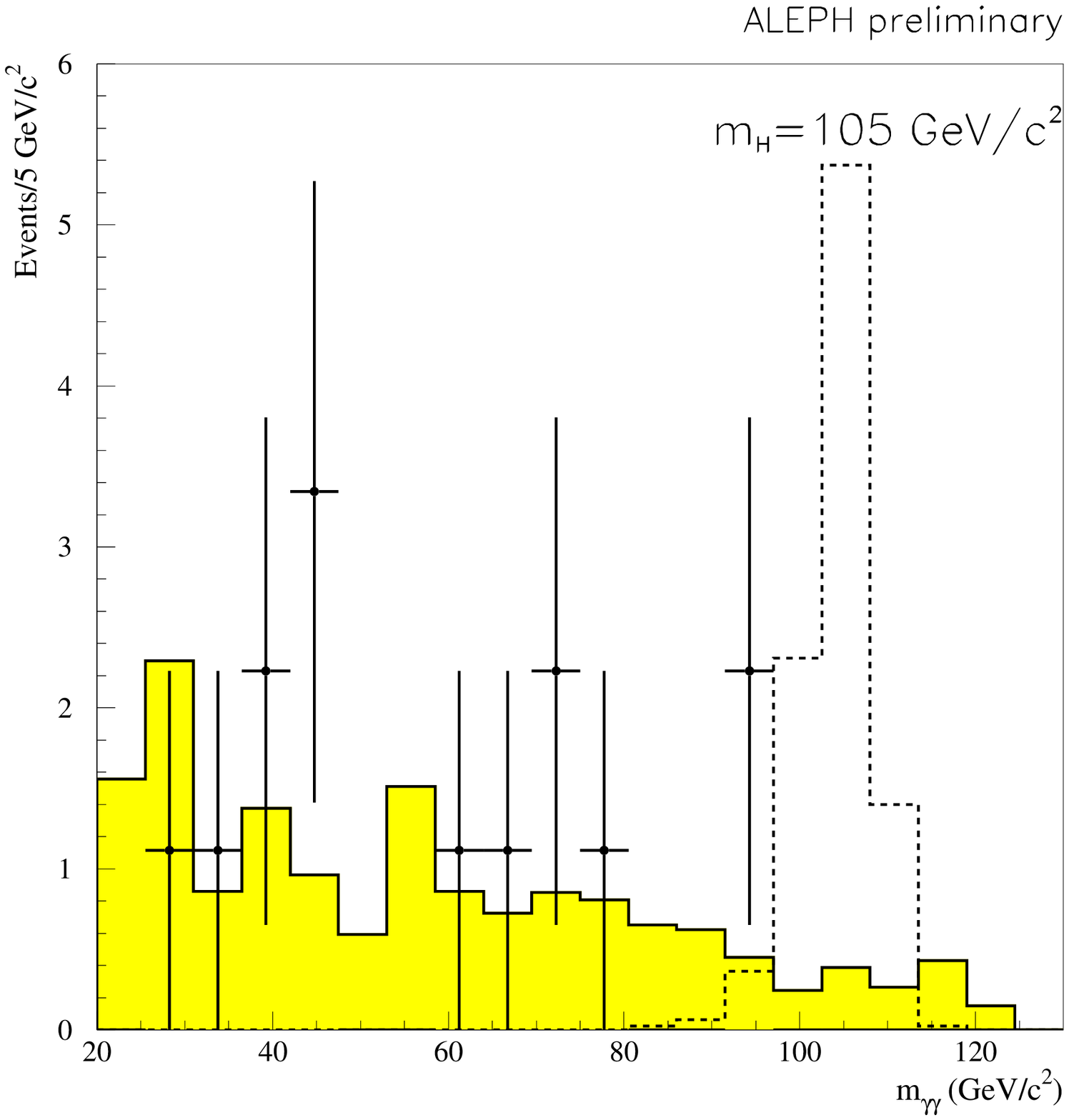}}  
\put(-6.2,5.8){(a)}& 
\hspace{0.cm}\mbox{\epsfxsize=.45\hsize\epsfbox{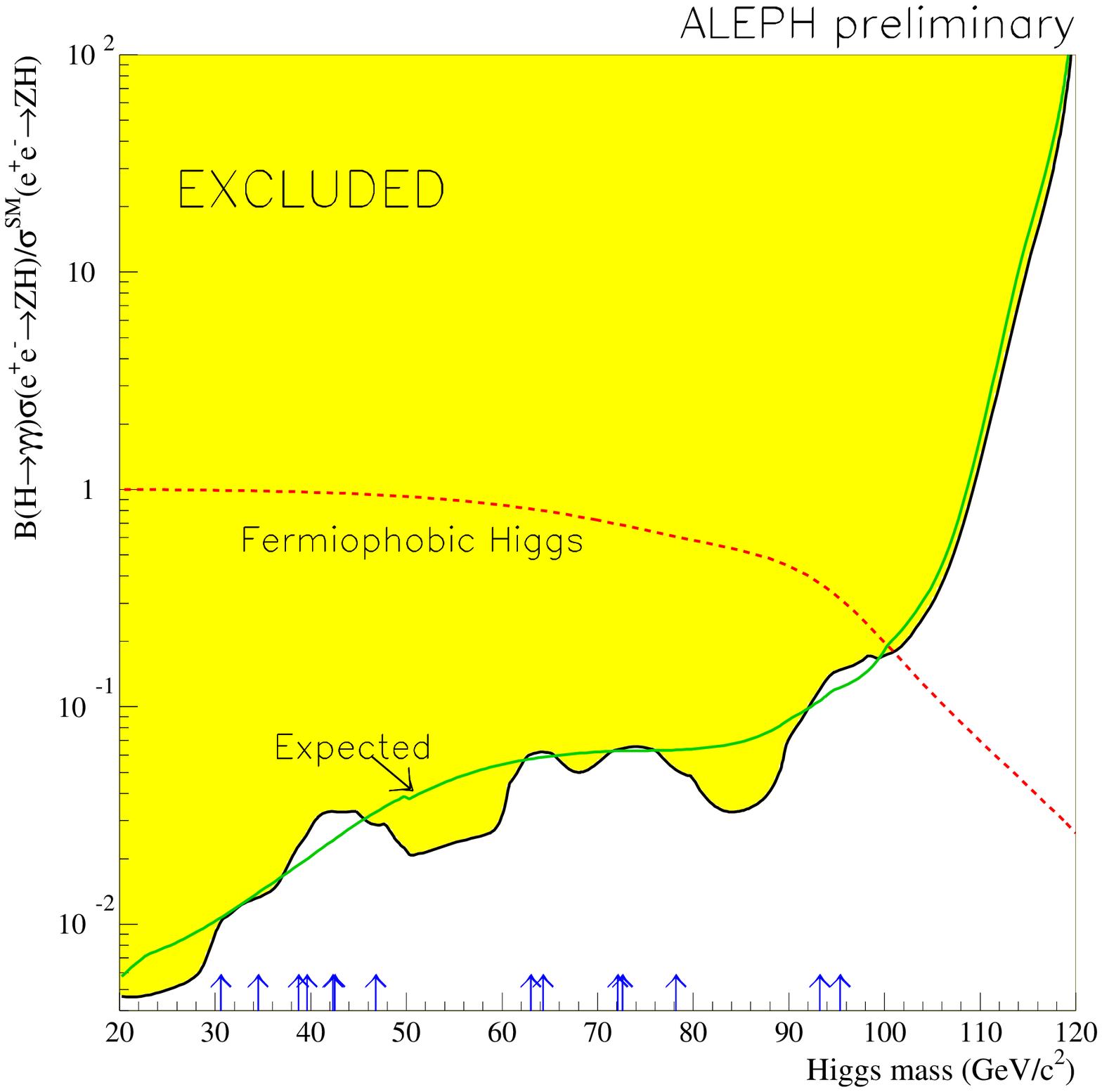}}
\put(-1.5,1.3){(b)} 
\end{tabular}
\caption{(a) Reconstructed $\gamma\gamma$ mass distribution for all 
data collected by ALEPH (dots), for the expected background 
(shaded histogram) and for a 105~\Gcs\ fermiophobic Higgs boson
signal (dashed histogram). (b) 95\% CL exclusion domain of 
the production rate of a Higgs boson decaying to two photons 
and produced via the Higgs-strahlung process normalised to
the standard model Higgs boson production cross section as a function
of the Higgs boson mass hypothesis (the reconstructed masses
of the candidate events are indicated by arrows for the purpose of 
illustration). 
\label{fermio}}
\end{center}
\end{figure}

In 2HDM of type I when the CP-even Higgs bosons do not mix, 
the lightest CP even Higgs 
boson does not couple to fermions. In a Higgs boson mass range well 
below 2\mW\ it predominantly decays to
a pair of photons. The topologies arising from such models are 
similar to those of the 
minimal standard model Higgs boson searches where a pair of photons is 
substituted for the pair of b-quark jets. Such topologies were 
searched for by all
four experiments. No evidence for a resonant 
production of a pair of photons was found
as illustrated in Fig.~\ref{fermio}a.
The negative result of these searches can be interpreted in a 
quasi-model-independent
fashion as an upper limit on the production 
rate of a Higgs boson decaying to a pair 
of photons normalised to the standard model Higgs boson production, as a 
function of the Higgs boson mass hypothesis. 
The ALEPH exclusion~\cite{fermiophobic1} is illustrated
in Fig.~\ref{fermio}b. The intercept of the model independent exclusion
and the branching fraction of the fermiophobic Higgs boson 
into a pair of photons,
which starts dropping severely above 100~\Gcs\ due to the growing 
contribution of $H\rightarrow WW^*$, gives the 
lower mass limit of the fermiophobic Higgs boson. 
A fermiophobic Higgs boson lighter than 101.0~\Gcs\ is excluded by ALEPH
with a sensitivity
of 100.2~\Gcs. The other experiments~\cite{fermiophobic2,opal,anomal} 
find similar results
---DELPHI, L3 and OPAL exclude fermiophobic Higgs boson masses below
98.0~\Gcs, 98.8~\Gcs\ and 97.8~\Gcs\ respectively.

%~\footnote{98.0~\Gcs,
%98.8~\Gcs\ and 97.8~\Gcs\ are the limits set by DELPHI, L3 and OPAL 
%respectively.}.

\subsection{2HDM, Charged Higgs bosons}

\begin{figure}[h]
\begin{center}
\begin{tabular}{cc}
\mbox{\epsfxsize=.45\hsize\epsfbox{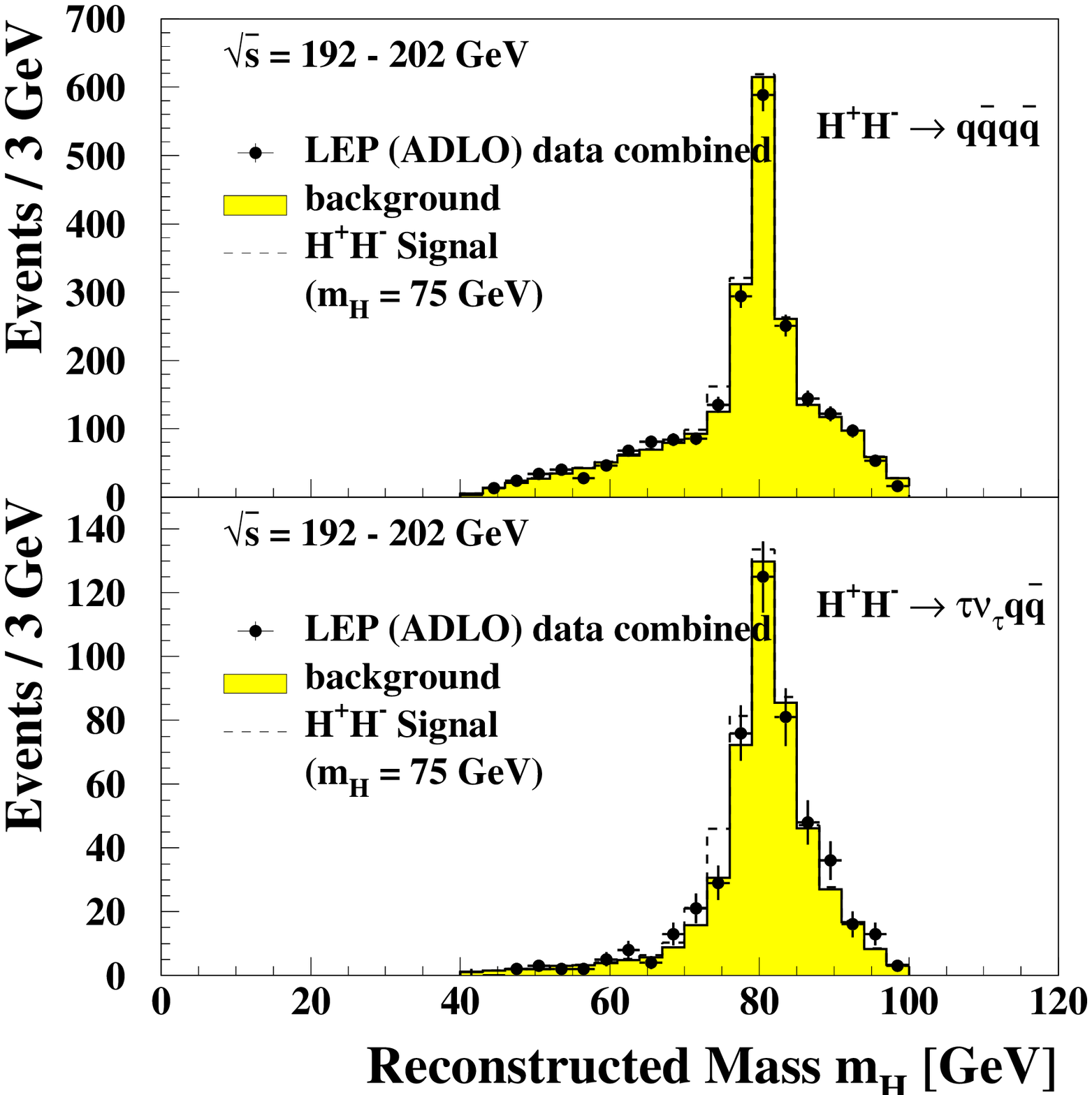}}
\put(-1.5,1.5){(a)}& 
\mbox{\epsfxsize=.45\hsize\epsfbox{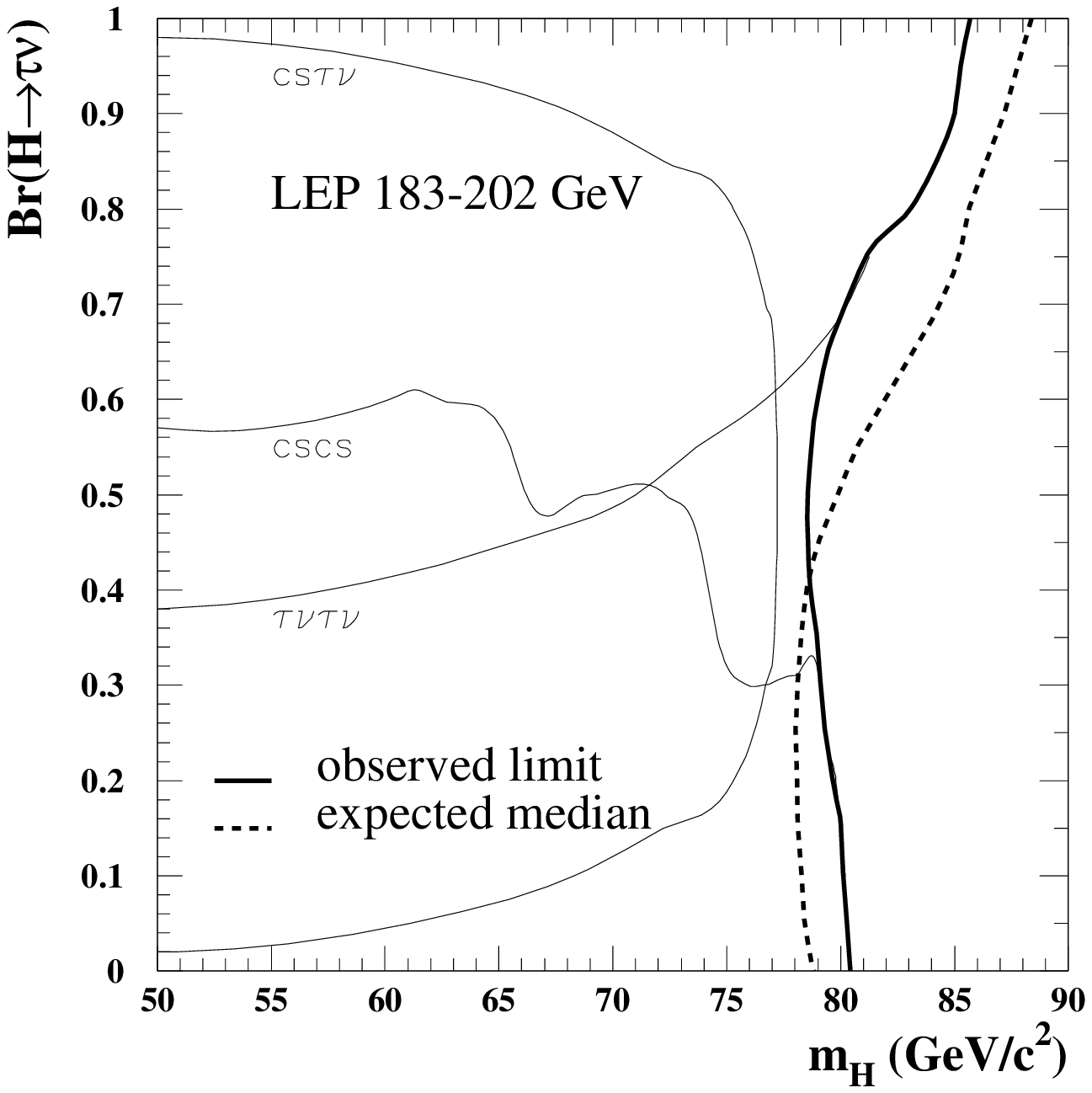}}  
\put(-1.9,1.5){(b)} 
\end{tabular}
\caption{(a) Reconstructed mass distribution for data (dots), 
for the expected background 
(shaded histogram) and for a 75~\Gcs\ charged Higgs bosons
signal (dashed histogram) in the fully hadronic (up) and 
semi-leptonic (down) channels. (b) 95\% CL exclusion domain of 
the charged Higgs boson branching to $\tau\nu$ as a function of their
mass.
\label{charged}}
\end{center}
\end{figure}

Charged Higgs bosons are produced at LEP via $\gamma$ and Z 
bosons exchange in the s-channel. The dominant decay modes are 
${\hbox{H}}^{\pm}\rightarrow\tau\nu_\tau,\hbox{cs}$. The relative 
contribution of these two modes depend
on the model parameter $\tan\beta$ (the ratio of the vev's of the two Higgs 
doublets). Analyses must therefore cover all possible final states arising
from charged Higgs bosons production, {\it i.e.}, $\tau^-\overline{\nu}_{\tau}
\tau^+\nu_{\tau}$, 
$\tau^-\overline{\nu}_{\tau}c\overline{s}$ and $c\overline{s}s\overline{c}$. 
Dedicated searches
for each of these topologies were performed by all LEP 
experiments~\cite{lephwg}.
No evidence for a signal was found, as illustrated in 
Fig.~\ref{charged}a for the fully hadronic and semi-leptonic channels. (The 
charged Higgs boson mass cannot be reconstructed in the fully leptonic
channel.) These results can be interpreted in terms of 95\% CL 
charged Higgs boson mass lower limits as a function of the 
leptonic branching fraction. As shown in Fig.~\ref{charged}b a 
lower charged Higgs 
boson mass limit on of 78.6~\Gcs\ with a sensitivity of 78.0~\Gcs\ can be 
set independently of the $\tau\nu_{\tau}$ branching fraction.

\section{Minimal Supersymmetric extension of the 
Standard Model (MSSM)}\label{subsec:mssm}

The presence of Higgs bosons in the standard model is quite unnatural, due to
the quadratically divergent contributions of the radiative corrections to 
their masses. Supersymmetry elegantly solves this problem and provides a 
framework for unified theories and quantum gravity.

\begin{wrapfigure}{r}{6cm}
\vspace{-.5cm}
\mbox{\epsfxsize=1.\hsize\epsfbox{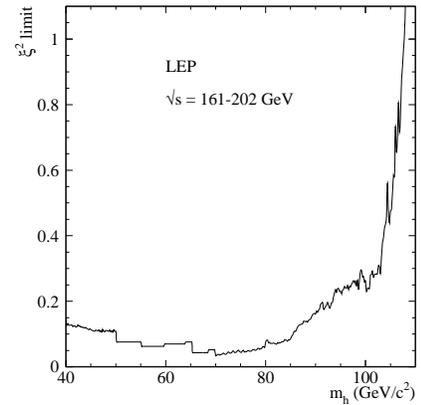}}
\caption{95\% CL Exclusion in the plane (\mh,$\xi^2$), where $\xi^2$ is the
the Higgs-strahlung production cross section normalised 
to that of the standard model. Here the Higgs boson is 
assumed to decay with with MSM couplings. 
\label{modindep}}
\vspace{-.5cm}
\end{wrapfigure}

In its minimal
version, the Higgs sector of the supersymmetric standard model 
is a type II
2HDM. Supersymmetry constrains Higgs bosons masses. 
At tree level $\mHpm \geq \mW$ and $\mh\leq \mZ |\cos 2\beta|$.
In the MSSM, two free parameters, 
commonly chosen either as \mA\ and $\tan\beta$ or \mh\ and $\tan\beta$, 
are required 
at tree level to describe the Higgs sector. However radiative corrections
due to the large Yukawa coupling to the top quark can increase the
theoretical upper bound on the lightest CP-even Higgs boson up to
about 130~\Gcs.
All the parameters relevant to the 
radiative corrections to the Higgs boson masses are chosen to
maximise \mh\ (\mh-max scenario). 

The Higgs-strahlung production of h is 
reduced, compared to the standard model, by a factor 
$\sin^2(\beta-\alpha)$. The results of searches for the standard model Higgs 
boson can be
turned into an 95\% CL exclusion domain in the  (\mh,$\sin^2(\beta-\alpha)$)
plane as illustrated in Fig.~\ref{modindep}.

\begin{figure}[h]
\begin{center}
\begin{tabular}{cc}
\mbox{\epsfxsize=.5\hsize\epsfbox{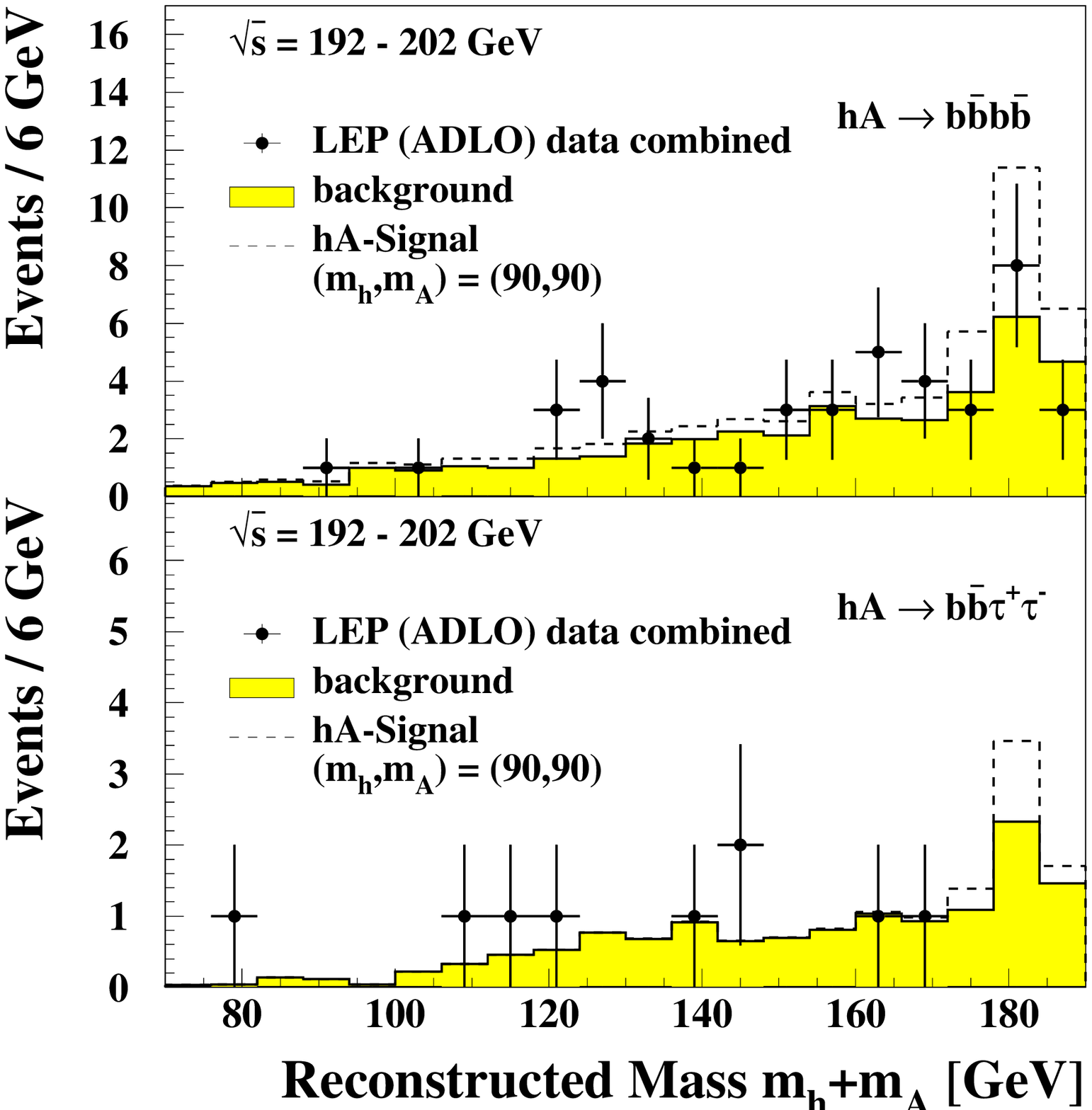}}
\put(-5.9,1.5){(a)}& 
\hspace{-.5cm}\mbox{\epsfxsize=.5\hsize\epsfbox{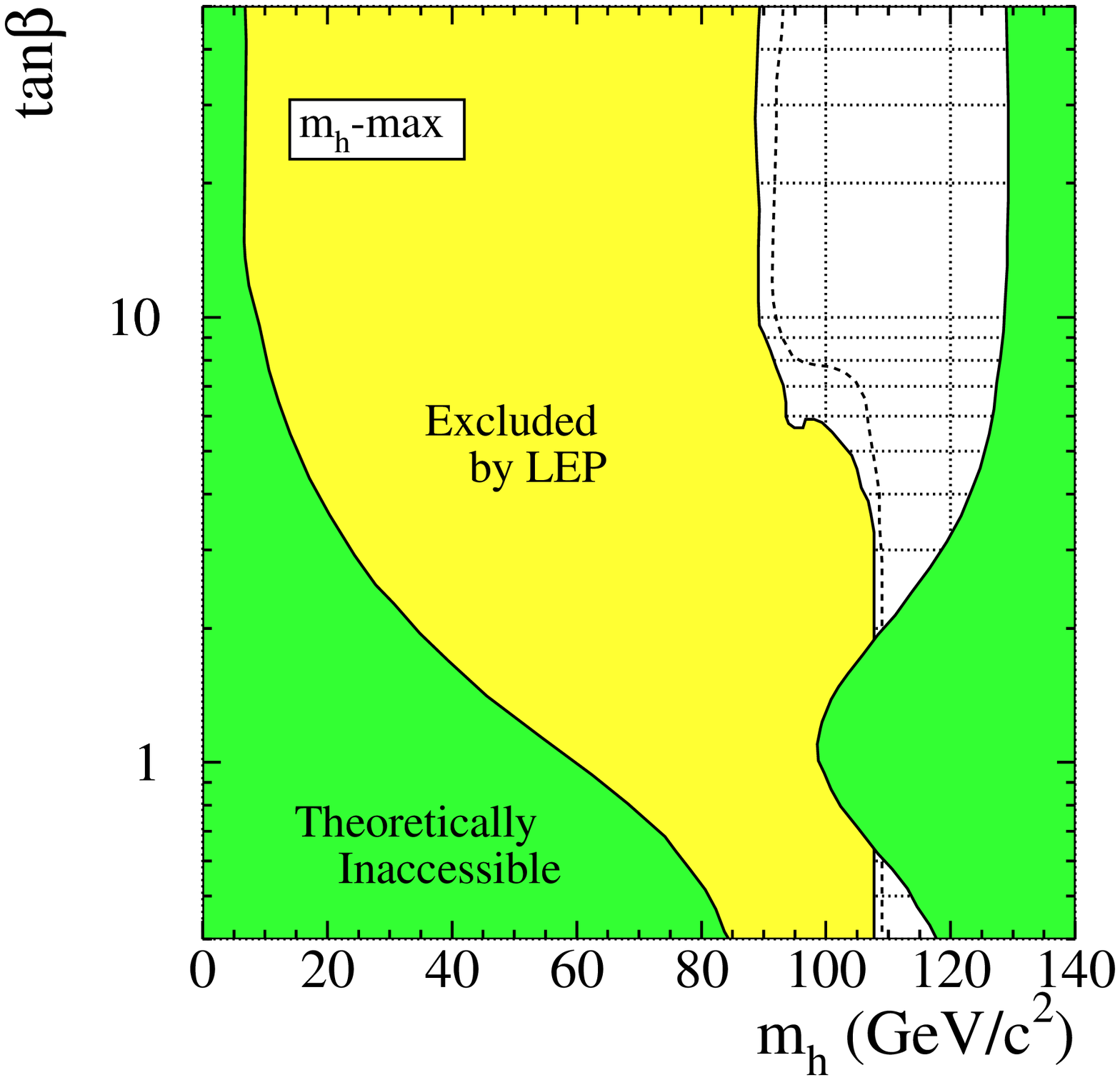}}
\put(-1.6,1.5){(b)} 
\end{tabular}
\caption{(a) Reconstructed mass sum \mh+\mA\ distribution for all 
data collected at LEP (dots), for the expected background 
(shaded histogram) and for a \mh=\mA=90~\Gcs\
signal (dashed histogram). (b) 95\% CL observed (light shaded) and expected
(dashed line) exclusion domains in the 
(\mh,$\tan\beta$) plane.
\label{mssm:mass}}
\end{center}
\end{figure}

For small
values of $\sin^2(\beta-\alpha)$ the Higgs-strahlung 
contribution vanishes but the
associated ${\rm e}^+{\rm e}^-\rightarrow {\rm hA}$ production process,
which cross section is proportional to $\cos^2(\beta-\alpha)$, 
becomes dominant.
In this regime, the h and A Higgs bosons are roughly mass
degenerate and both decay predominantly to a pair of b-quarks. This 
process is thus searched for in two additional topologies~\cite{lephwg}. 
1) Four jets likely
to originate from b quarks; and 2) Two b-like jets and a pair of taus. 
No evidence for a signal was found 
as illustrated in Fig.~\ref{mssm:mass}a. The negative
result of these searches combined with the search for the standard model 
Higgs boson can be interpreted in terms of an exclusion domain in the 
(\mh,$\tan\beta$) plane as shown in Fig.~\ref{mssm:mass}b. 
Neutral Higgs boson masses in excess of 88.3~\Gcs\ are excluded at the 
95\% CL with a sensitivity of 90.5~\Gcs. 
The lower mass limit on the 
CP-odd Higgs boson is set to 88.4~\Gcs\ 
with
a sensitivity of 91.1~\Gcs. These limits have proven 
to be robust while
scanning over all parameters pertaining to radiative corrections to the 
Higgs boson masses~\cite{scan1,scan2,scan3}. 
As can be seen in Fig.~\ref{mssm:mass}b 
values of $\tan\beta$ comprised within [0.7,1.8] 
are excluded at the 95\% CL, for m$_{{\hbox{\scriptsize top}}}$=174.3~\Gcs\ and [0.8,1.5] 
for m$_{{\hbox{\scriptsize top}}}$=179.4~\Gcs. 

\section{Further Exotism}

\subsection{Invisible decays of Higgs bosons}

\begin{figure}[h]
\begin{center}
\begin{tabular}{cc}
\mbox{\epsfxsize=.45\hsize\epsfbox{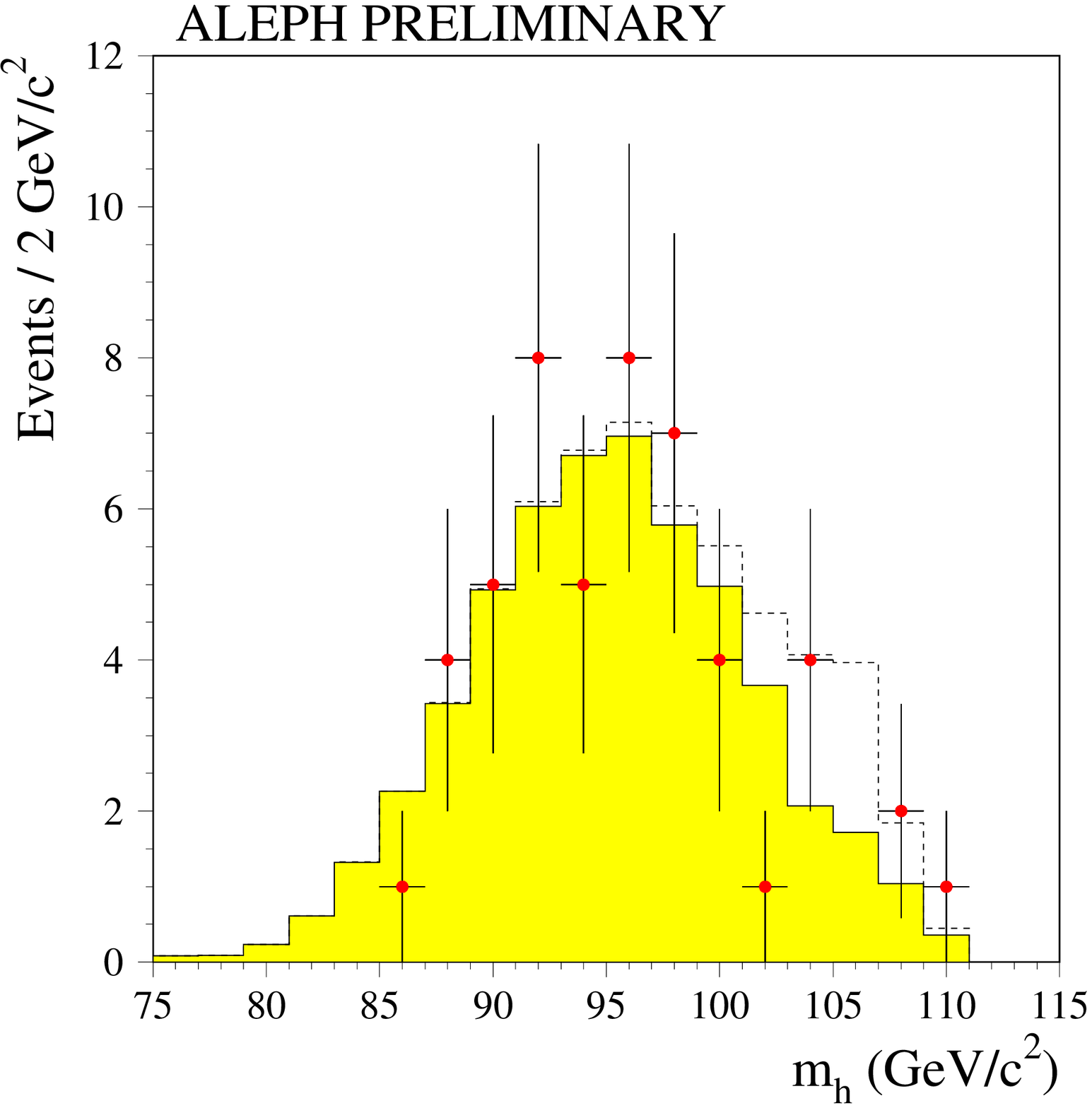}}  
\put(-5.5,5.55){(a)}& 
\hspace{0.cm}\mbox{\epsfxsize=.45\hsize\epsfbox{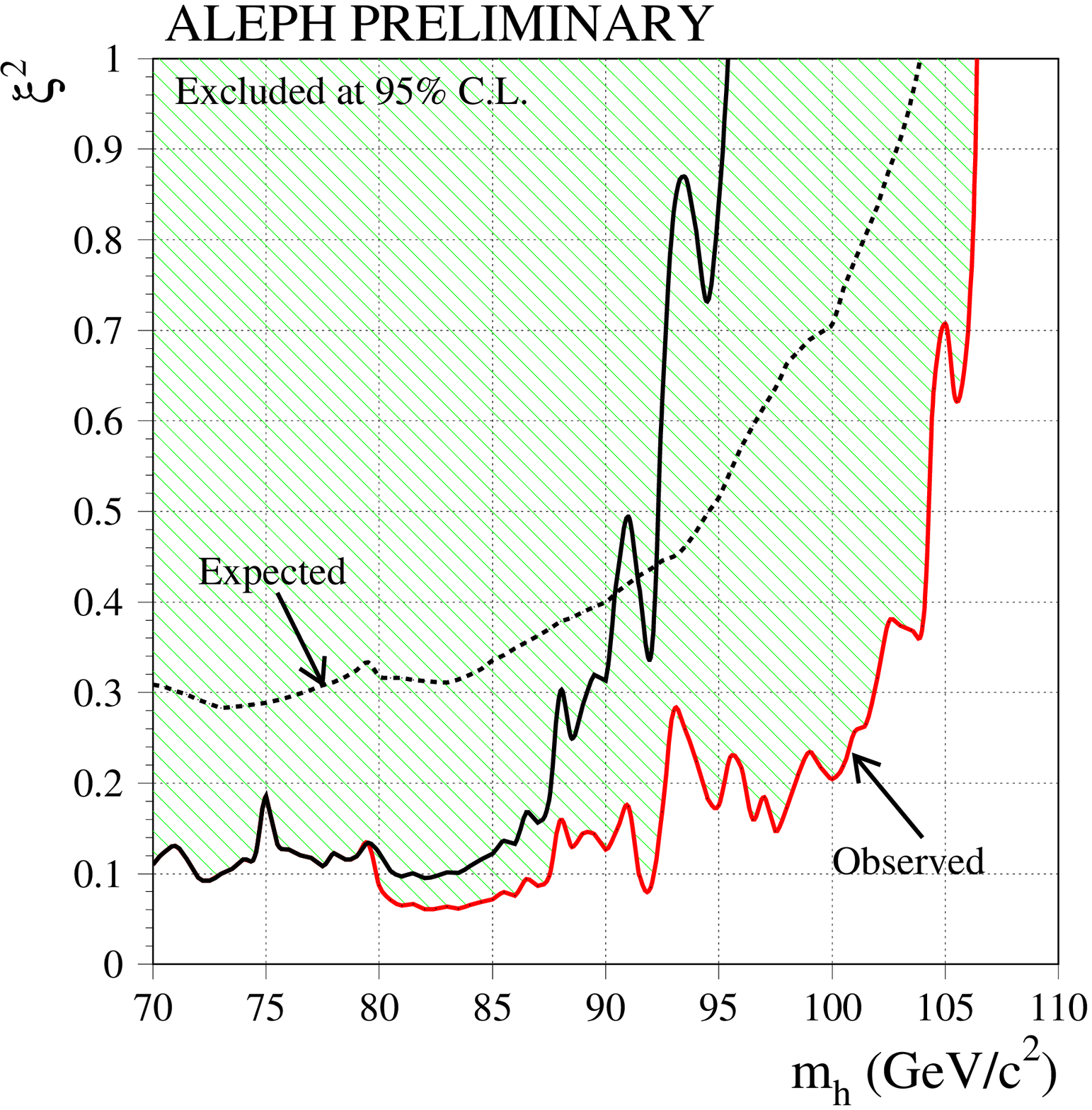}}
\put(-5.5,5.55){(b)} 
\end{tabular}
\caption{(a) Reconstructed mass distribution for all 
data collected with the ALEPH detector (dots), for the expected 
background (shaded histogram) and for a 
105~\Gcs\ invisible Higgs boson signal (dashed histogram). 
(b) 95\% CL exclusion domain of 
the production rate of an invisible Higgs boson produced via 
the Higgs-strahlung process normalised to
the standard model higgs boson production cross section as a function
of the Higgs boson mass hypothesis.
\label{invisible}}
\end{center}
\end{figure}

In a MSSM conserving R-parity, the domains in which Higgs bosons
 can decay to
neutralinos (supersymmetric partners of the neutral Higgs and gauge
bosons, here the lightest neutralino is assumed to be the lightest 
supersymmetric particle) is tightly constrained by direct searches 
for charginos (supersymmetric 
partners of charged Higgs and gauge bosons). Nevertheless the latter
exclusion assumes universality of gaugino masses at a 
large unification scale. If this assumption
is relaxed the indirect chargino constraints do not necessarily 
hold and 
the Higgs boson can thus decay to a pair of neutralinos. 
Neutralinos 
are weakly interacting neutral particles, like neutrinos. They therefore
escape detection and the Higgs boson decays invisibly. 
A variety of other 
theories, from models involving Majorons to 
large extra dimensions, predict such decays.

The two 
main topologies under consideration correspond to the Z decays to a pair
of quarks and to a pair of leptons (either an electron or a muon). 1)
Two acoplanar jets; and 2) two leptons, both with missing
energy. Such topologies were 
searched for by all four LEP collaborations. No evidence for a signal was
found, as shown in Fig.~\ref{invisible}a for ALEPH data 
only~\cite{invisible1}. This
result can then be interpreted in a quasi-model-independent fashion  
as a limit on the production rate of an invisible Higgs boson
normalised to the standard model cross section:
$\xi^2=Br({\rm h}\rightarrow{\rm invisible})
\times(\sigma_{hZ}/\sigma_{HZ}^{\rm MSM})$ as a function of \mh\ as 
illustrated in Fig.~\ref{invisible}b. Higgs 
bosons decaying invisibly and produced at standard model rate ($\xi^2=1$)
with masses below 106.4~\Gcs\ are excluded at 95\% CL with a 
sensitivity of 103.9~\Gcs. The other experiments find similar 
results ---93.8~\Gcs,
100.5~\Gcs\ and 94.4~\Gcs\ are the limits (for $\xi^2=1$) 
set by DELPHI (with data up to 189~GeV), 
L3 and OPAL respectively---
%~\footnote{93.8~\Gcs,
%100.5~\Gcs\ and 94.4~\Gcs\ are the limits set by DELPHI (up to 189~GeV), 
%L3 and OPAL respectively for $\xi^2=1$.} 
(unless their high energy data set had not yet been
entirely analysed)~\cite{opal,invisible1,invisible2,invisible3}.

\subsection{Anomalous Higgs couplings}                  

The effect of theories which supersede 
the standard model at some large scale $\Lambda$ can also be parametrised at
low energy by an effective Lagrangian in a model independent manner. 
The
simplest corrections to the standard model Lagrangian distorting the 
couplings of 
\begin{wrapfigure}{l}{7.5cm}  
\vspace{-.8cm}
\mbox{\epsfxsize=1.\hsize\epsfbox{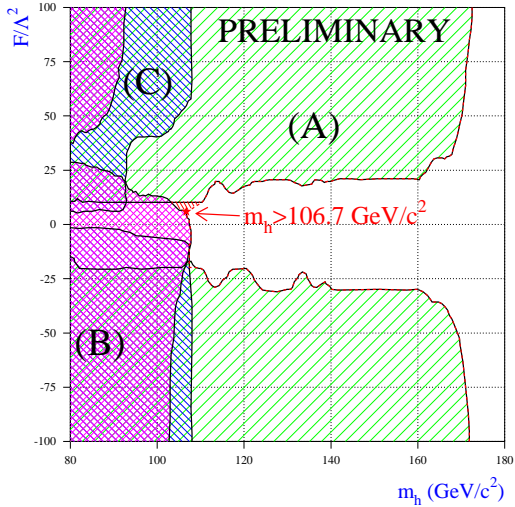}}
\vspace{-.8cm}
\caption{95\% CL limits on the generic anomalous coupling $F$
as a function of \mh. \label{anomalous}}
\vspace{-.5cm}
\end{wrapfigure}
Higgs bosons originate from terms of the type ${\mathcal L}_{\rm 
eff}=\sum_n (f_n/\Lambda^2){\mathcal O}_n$
where the ${\mathcal O}_n$ operator involves vector boson and/or Higgs boson 
fields with couplings $f_n$. 
Such terms can give rise to anomalous couplings
of the type $g_{{\hbox{\scriptsize {H}}}\gamma\gamma}$, 
$g_{{\hbox{\scriptsize {HZ}}}\gamma}$ 
and $g_{{\hbox{\scriptsize HZZ}}}$  which 
affect the expected phenomenology of a standard Higgs sector. For instance
the Higgs boson can be produced along with a photon and decay itself to a 
pair of photons. 
The DELPHI collaboration performed a searched 
for topologies with three photons in the framework of anomalous 
couplings of Higgs bosons~\cite{anomal}. 
No evidence for a signal was 
found and limits on the generic $F$ coupling ---here all
the underlying couplings $f_n$ relevant for the Higgs anomalous couplings are
assumed to be equal to $F$---  can be set as a function of \mh\
as shown in Fig.~\ref{anomalous} (region A).

\newpage

In this general framework, the Higgs boson can also predominantly 
couple to photons but still be produced via the Higgs-strahlung process 
and thus give rise to the topologies searched for in the framework of 
fermiophobicity. 
The ALEPH searches presented in Section~\ref{sec:fermio}
were reinterpreted here in terms of an exclusion of 
the coupling $F$ as a function of the Higgs boson mass 
hypothesis, as illustrated in Fig.~\ref{anomalous} (region C).

These searches require non-zero anomalous couplings  
and can therefore not explore the small $F$ region. 
However the searches for the 
standard model Higgs boson can also be reinterpreted in this more general 
framework as shown in Fig.~\ref{anomalous} (region B) where the ALEPH 
standard model Higgs boson searches are used. Finally a combination 
of all these 
searches allow a mass lower limit to be set at $106.7$~\Gcs\ 
on the mass of a Higgs boson 
allowing it to couple anomalously to photons with the assumption that all
relevant couplings $f_n$ are equal. 

\section{Summary and prospects}

All LEP experiments have made meaningful efforts to corner 
all possible forms in which Higgs bosons might appear. All the
topologies and the relevant theoretical frameworks which were investigated
with the data collected in 1999 have been presented. The resulting mass lower
limits on the mass of Higgs bosons in each of these models and 
all the topologies are sketched in Table~\ref{topologies}.

Topologies not covered in this review are those searched for 
in flavour-independent analyses performed by OPAL~\cite{flavindep}
(interpreted in general 2HDM models) and ALEPH~\cite{scan1} 
(to exclude pathological MSSM parameter sets in their 
scan beyond benchmark configurations).

\begin{table}[h]
\caption{Synoptic panorama of all topologies searched for in the data
collected in 1999 within the framework of Higgs boson 
searches. The models in which 
these topologies are relevant are also indicated.\label{topologies}}
\vspace{0.4cm}
\begin{center}
\begin{tabular}{|c|c|c||c|c|}
\hline \hline
\multicolumn{3}{|c||}{Minimal Standard Model (\mH$>107.7$~\Gcs)} &
\multicolumn{2}{|c|}{MSSM (\mh$>88.3$~\Gcs)}\\ \hline 
\mbox{\epsfxsize=.11\hsize\epsfbox{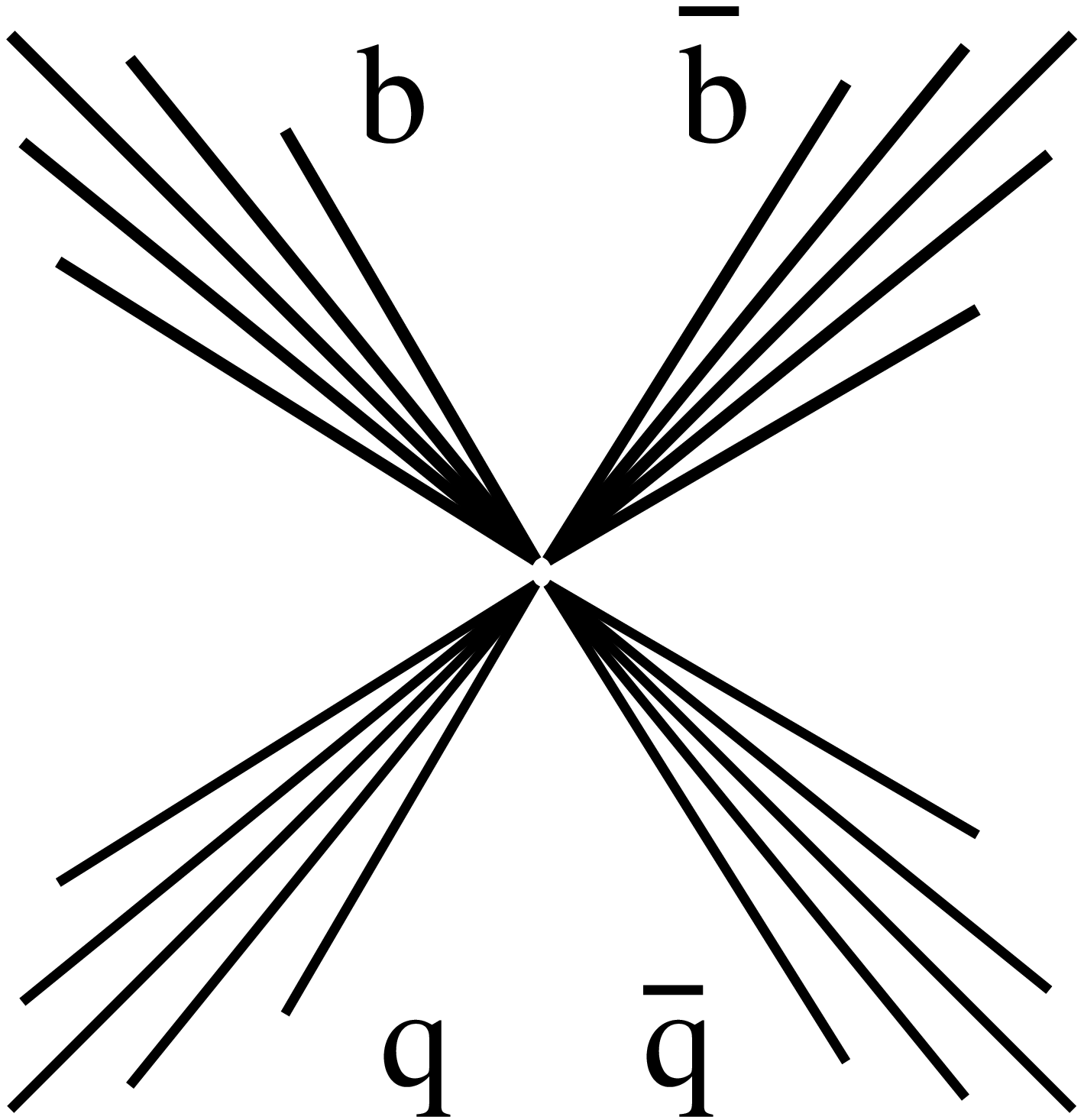}}& 
\mbox{\epsfxsize=.11\hsize\epsfbox{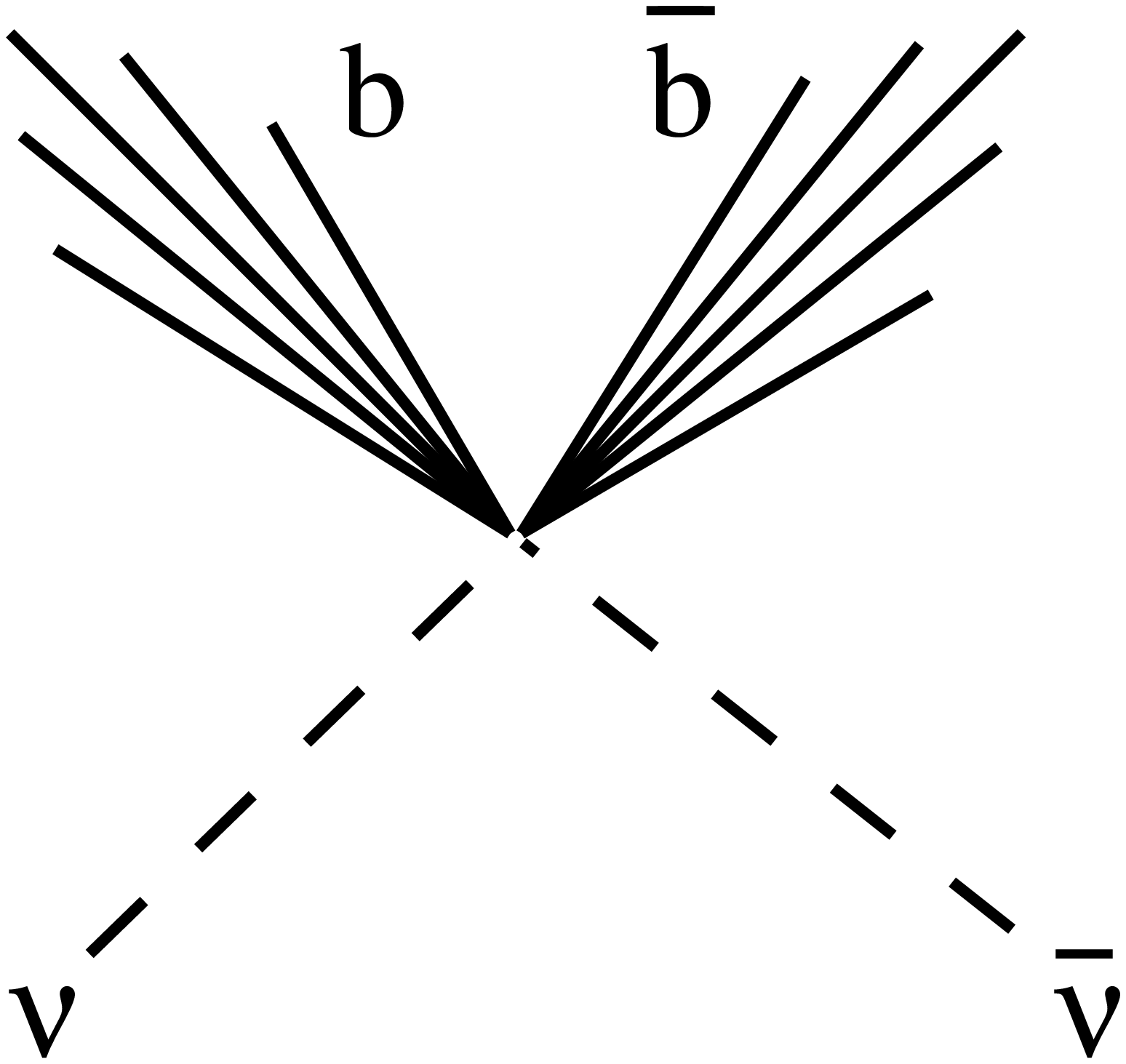}}&
\mbox{\epsfxsize=.11\hsize\epsfbox{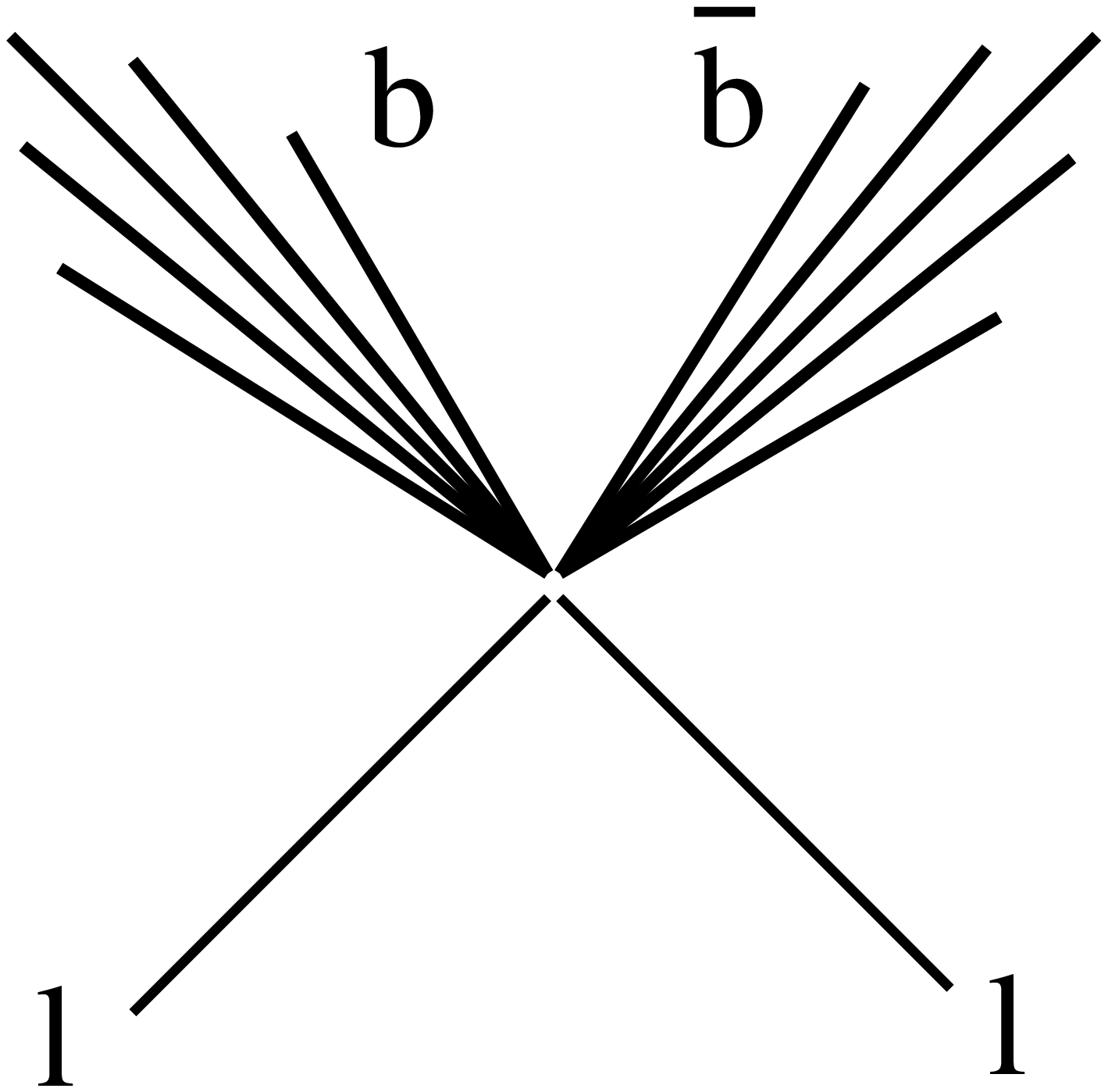}} &
\mbox{\epsfxsize=.11\hsize\epsfbox{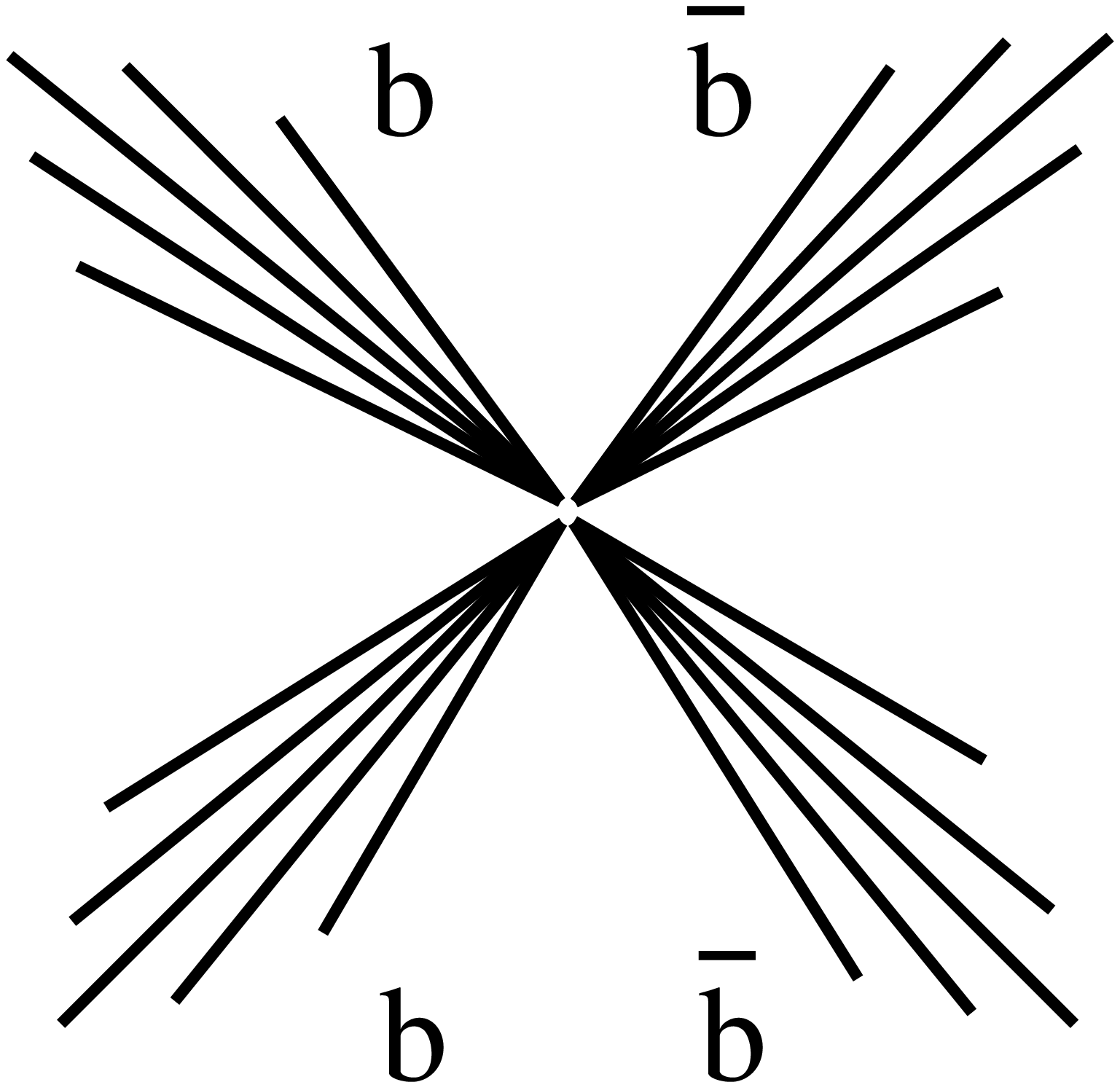}}& 
\mbox{\epsfxsize=.11\hsize\epsfbox{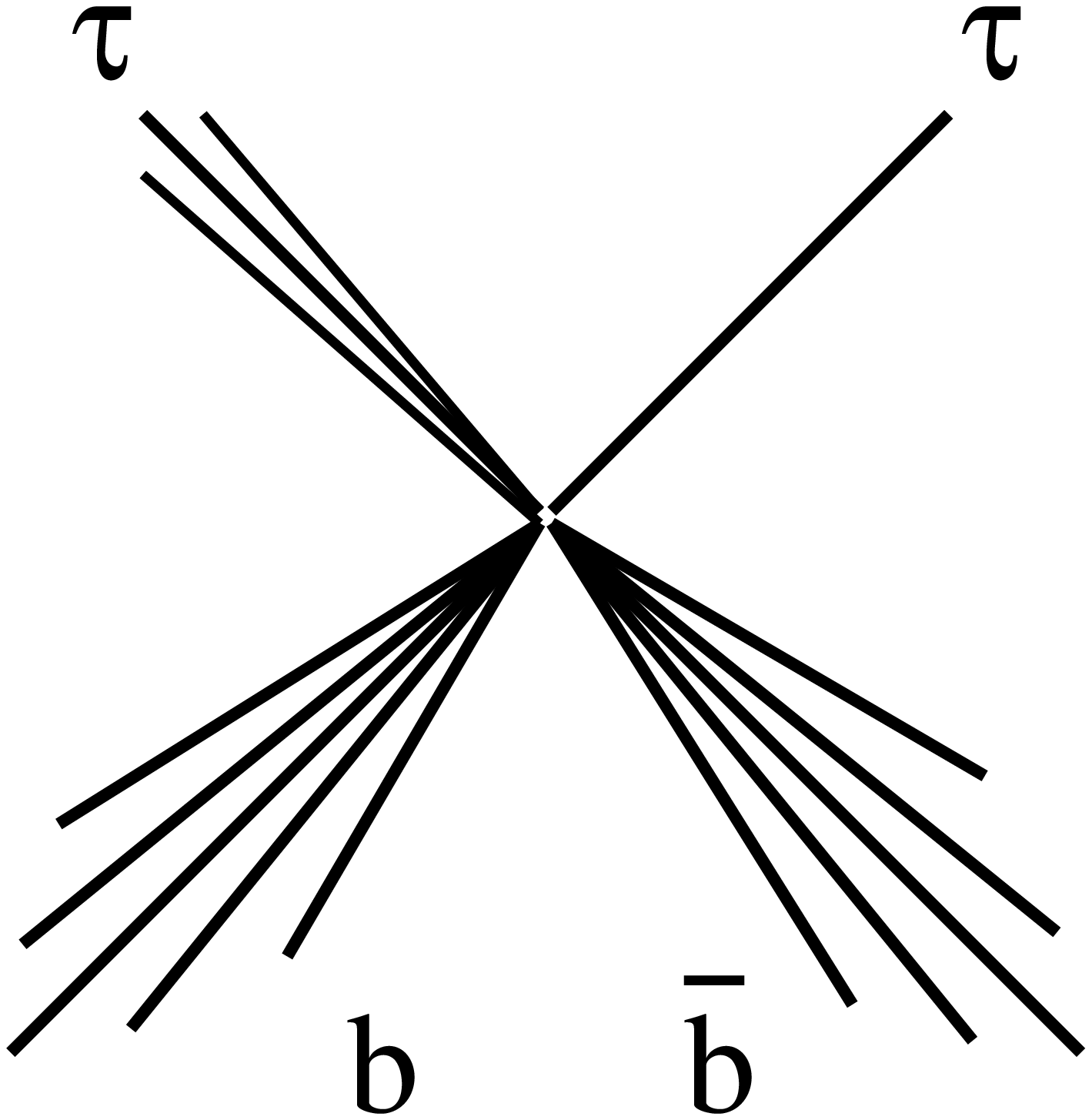}}\\ \hline
4 Jets & 2 Acoplanar jets & Leptons & 4 b-jets & Taus \\ 
&and missing energy & & & \\ \hline \hline
\multicolumn{3}{|c||}{2HDM Type I (\mh$>101.0$~\Gcs)} &
\multicolumn{2}{|c|}{Invisible decays (\mh$>106.4$~\Gcs)}\\ \hline 
\mbox{\epsfxsize=.11\hsize\epsfbox{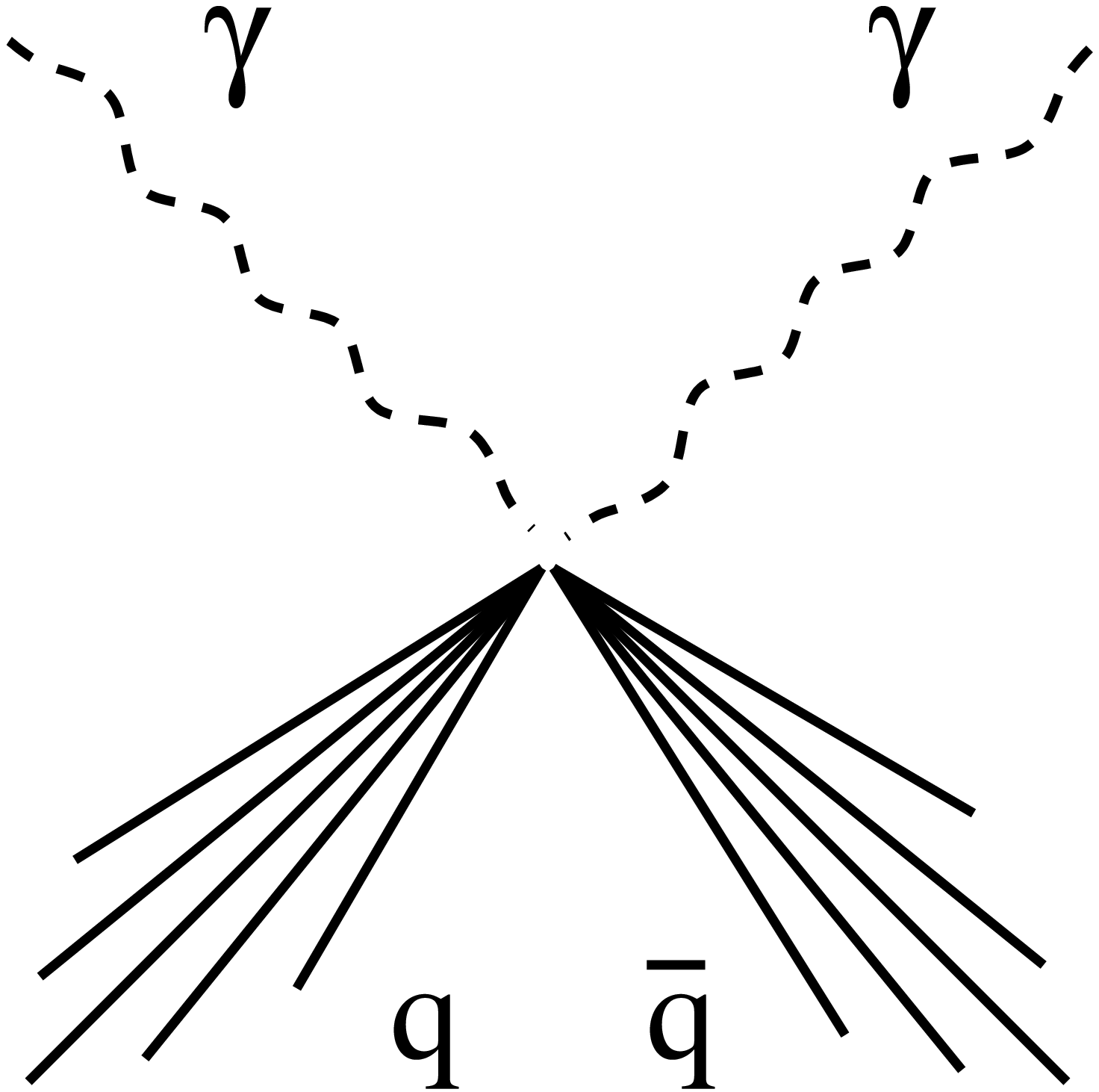}}& 
\mbox{\epsfxsize=.11\hsize\epsfbox{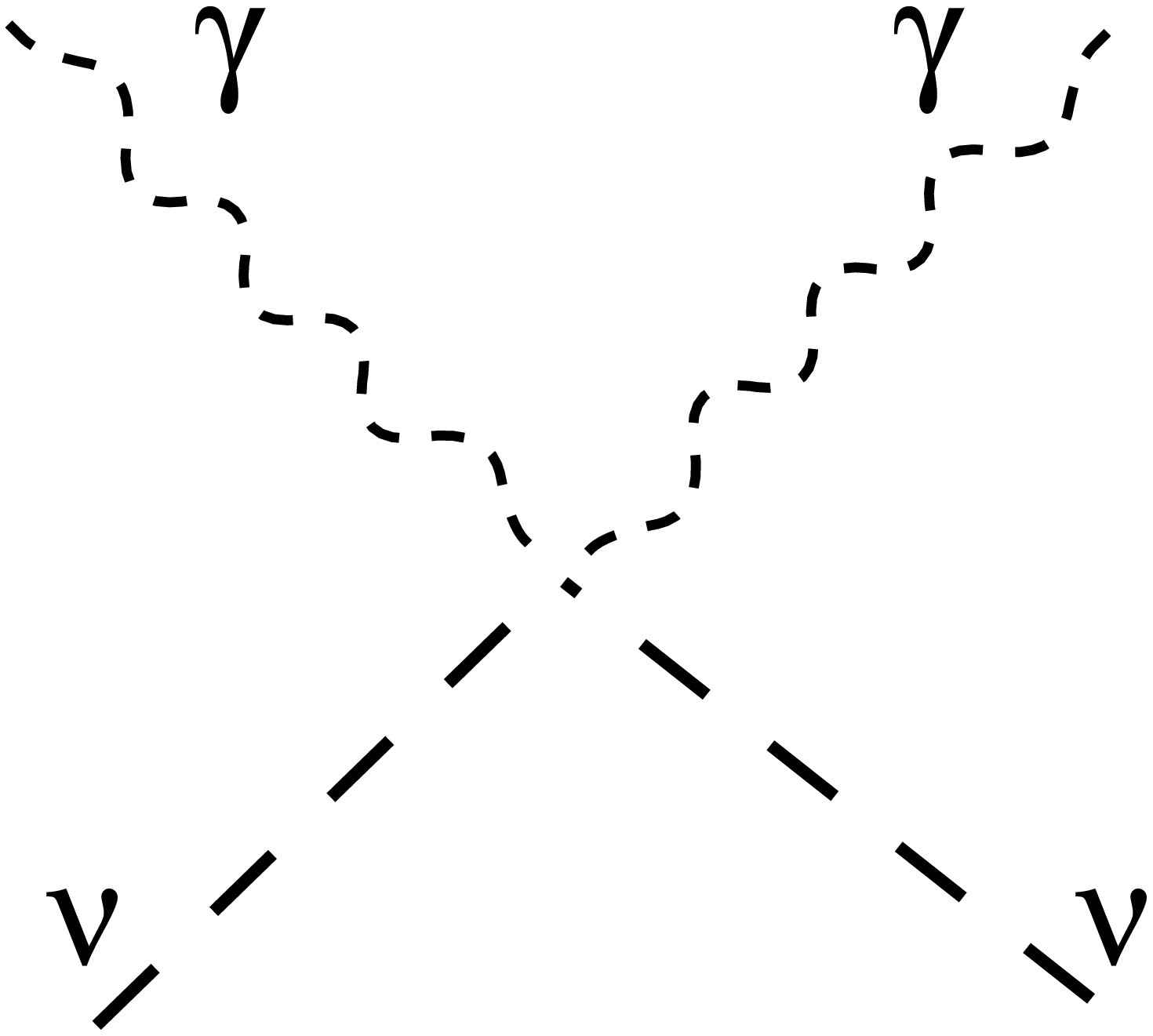}}&
\mbox{\epsfxsize=.11\hsize\epsfbox{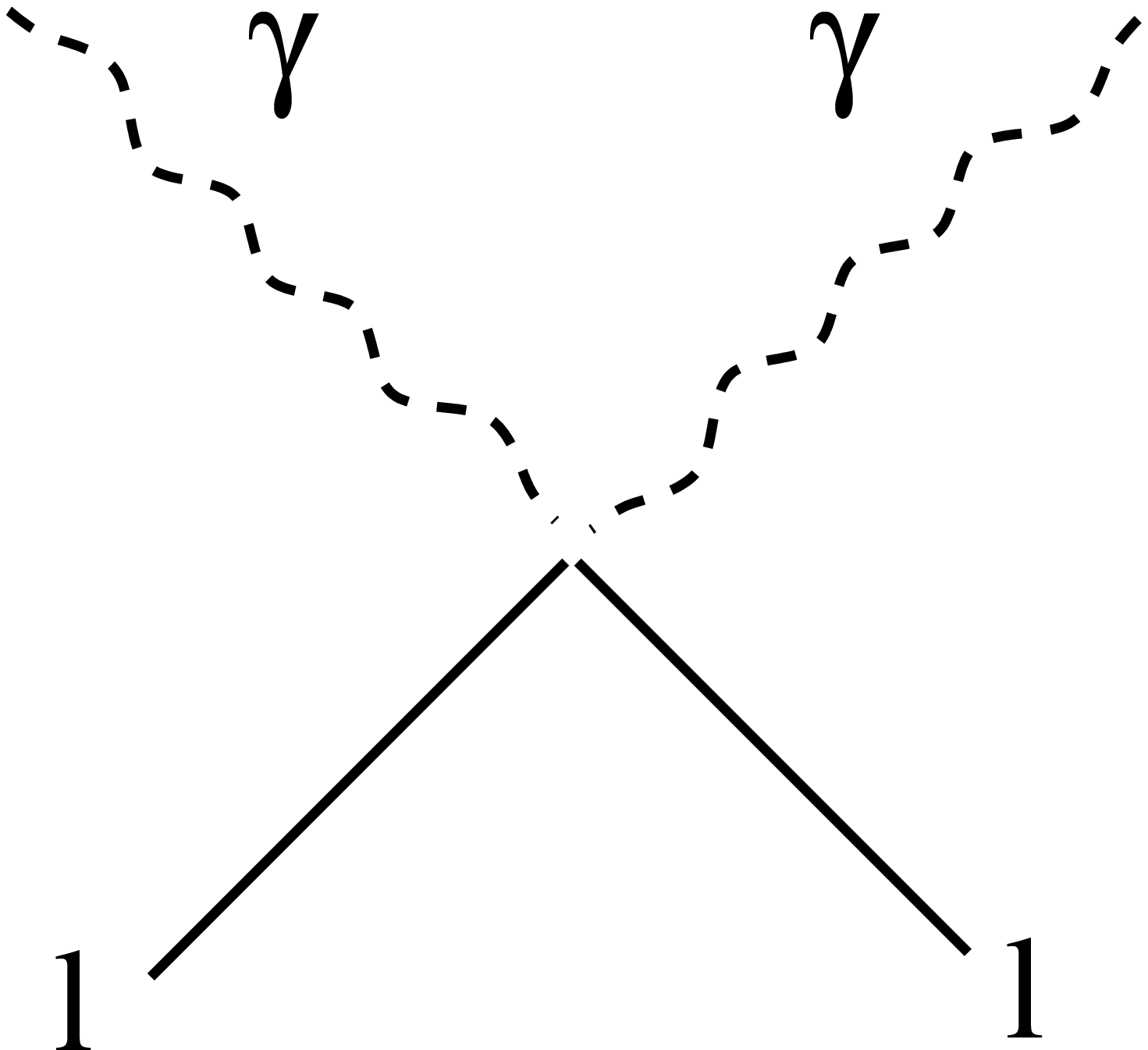}} &
\mbox{\epsfxsize=.11\hsize\epsfbox{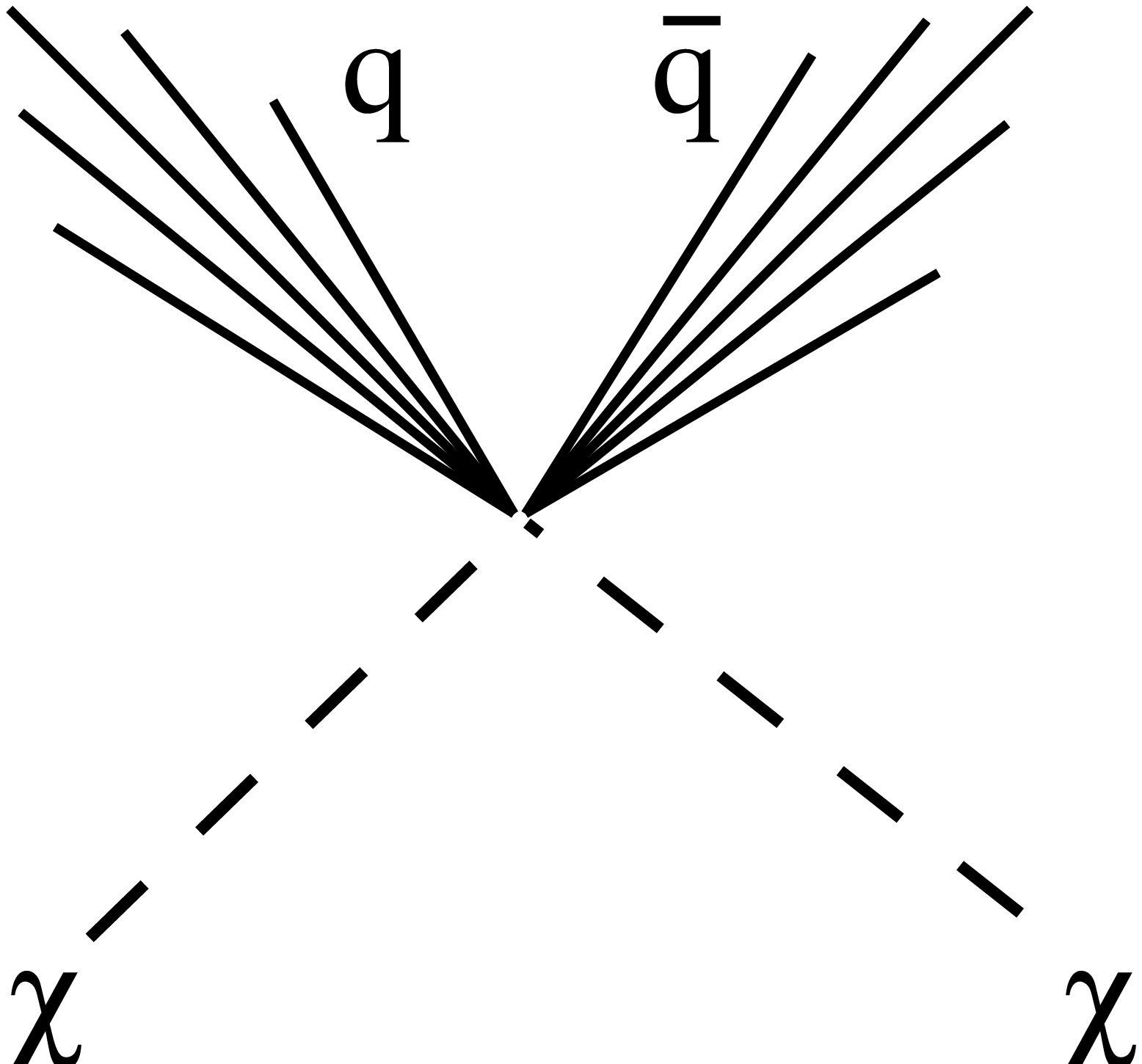}}&
\mbox{\epsfxsize=.11\hsize\epsfbox{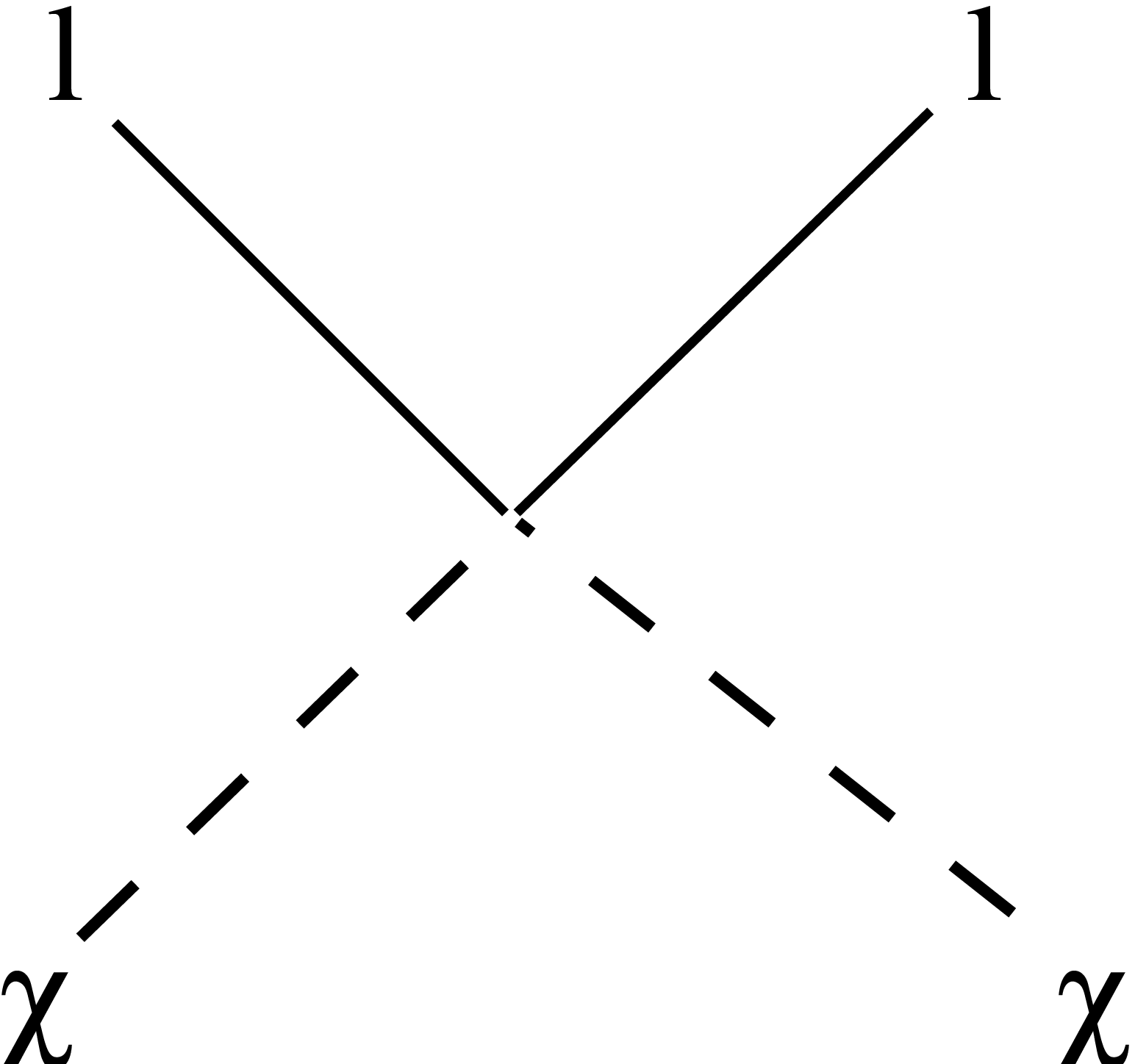}}\\ \hline
Hadronic & 2 Acoplanar photons & Leptonic & Acoplanar jets & Leptonic 
\\ \hline \hline
\multicolumn{3}{|c||}{2HDM type I \& II (\mHpm$>78.6$~\Gcs)} &
\multicolumn{2}{|c|}{Anomalous couplings (\mh$>106.7$~\Gcs)}\\ \hline
\mbox{\epsfxsize=.11\hsize\epsfbox{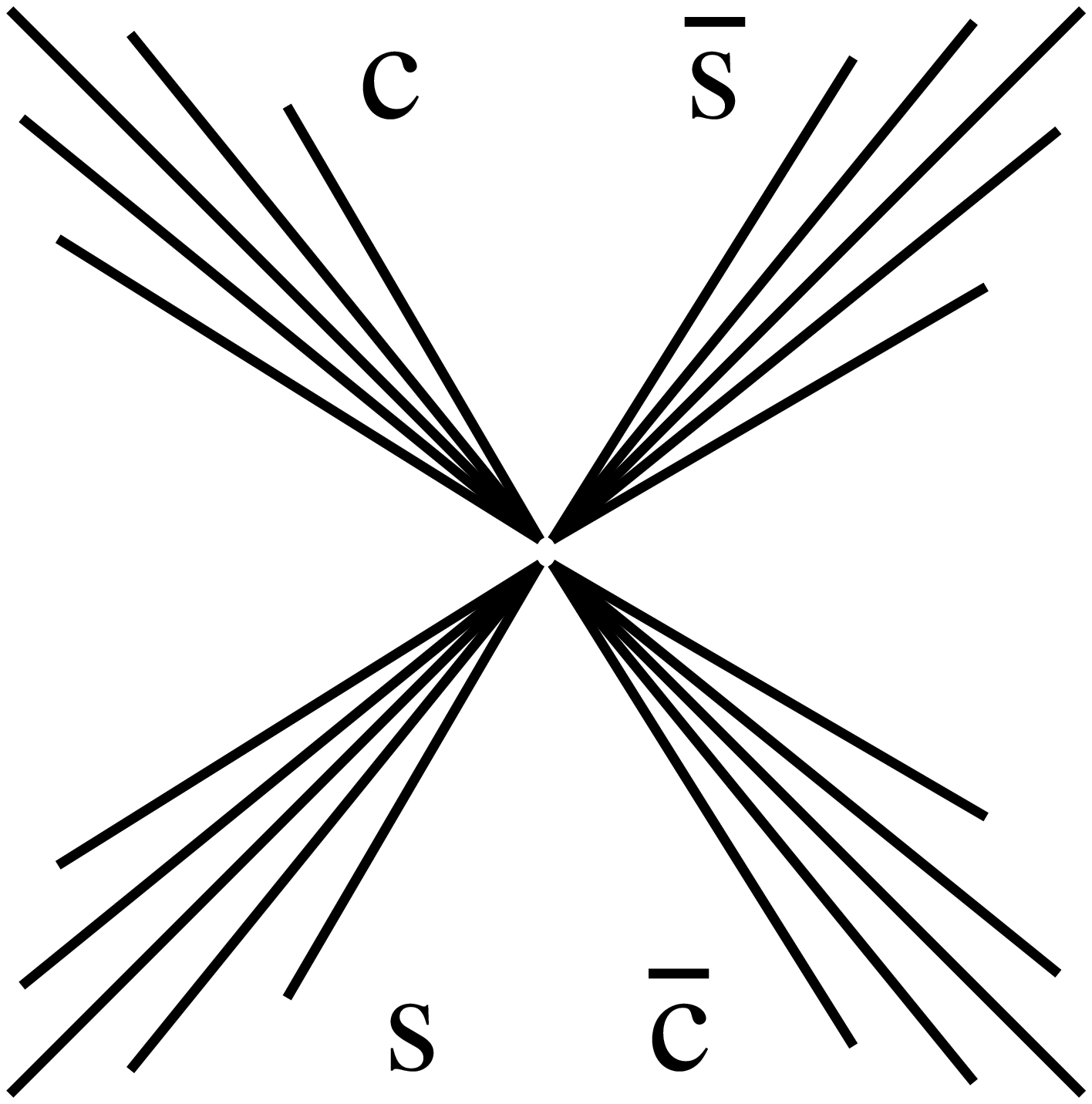}}&
\mbox{\epsfxsize=.11\hsize\epsfbox{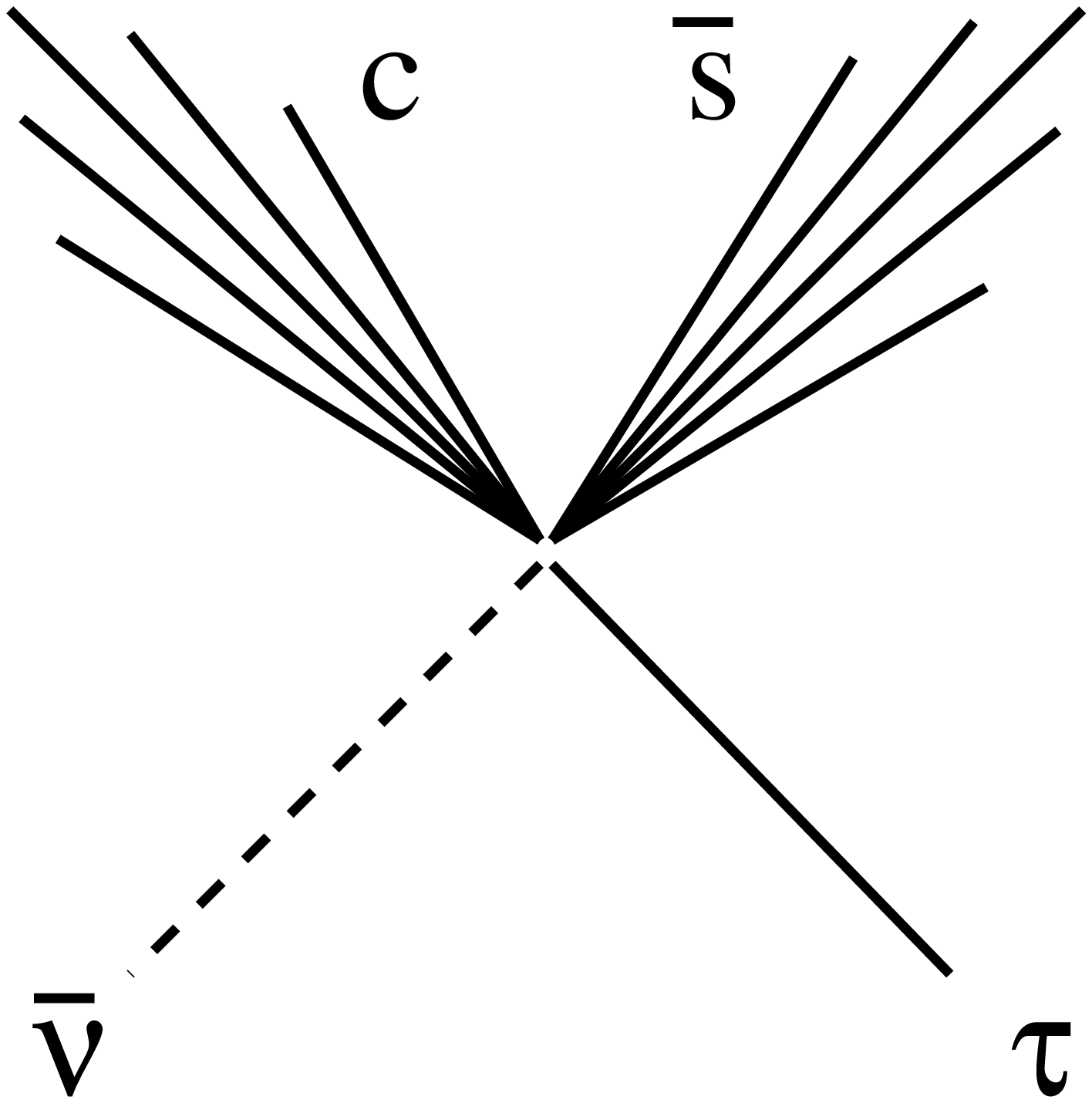}}&
\mbox{\epsfxsize=.11\hsize\epsfbox{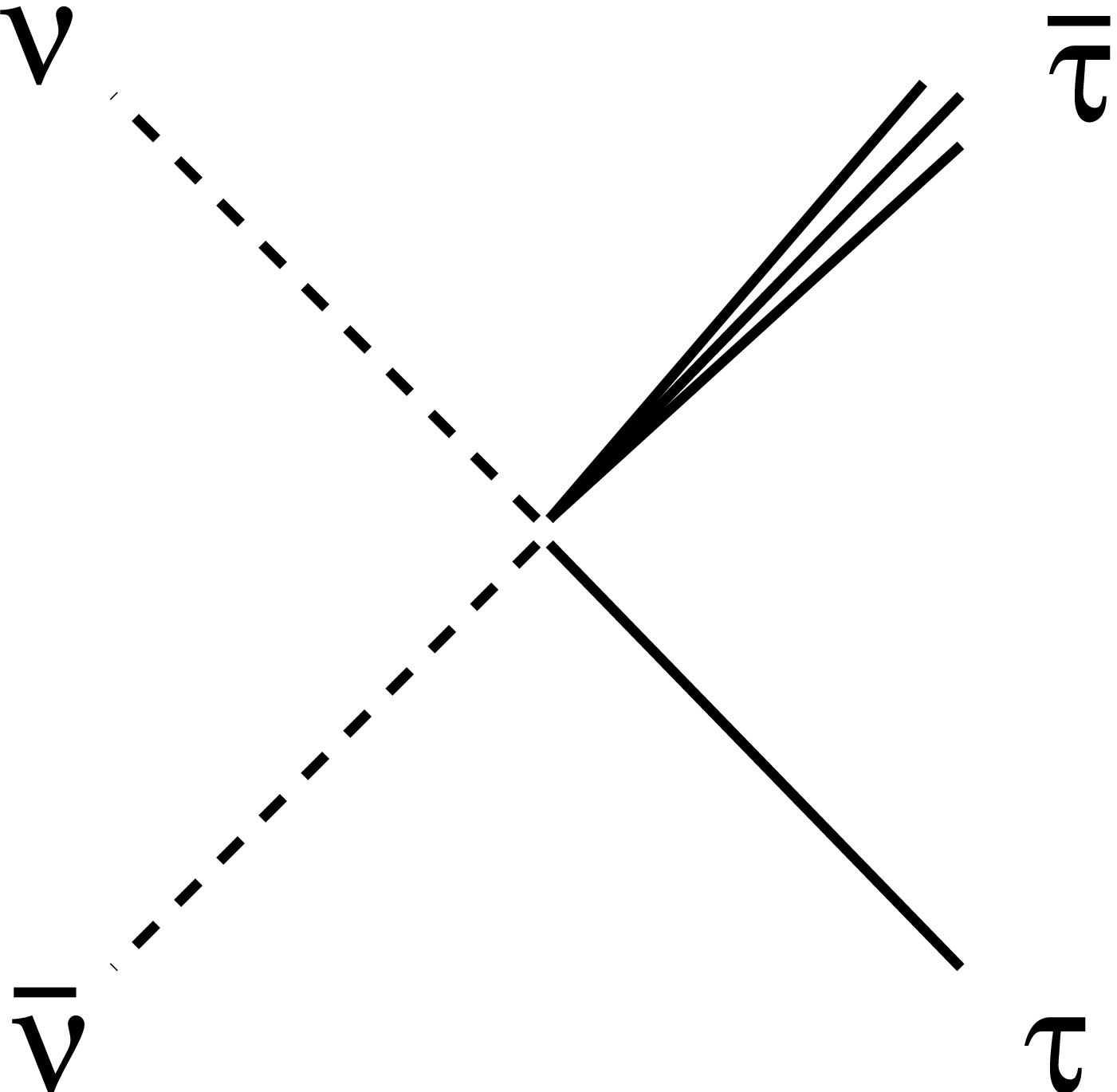}} &
\multicolumn{2}{|c|}{
\mbox{\epsfxsize=.11\hsize\epsfbox{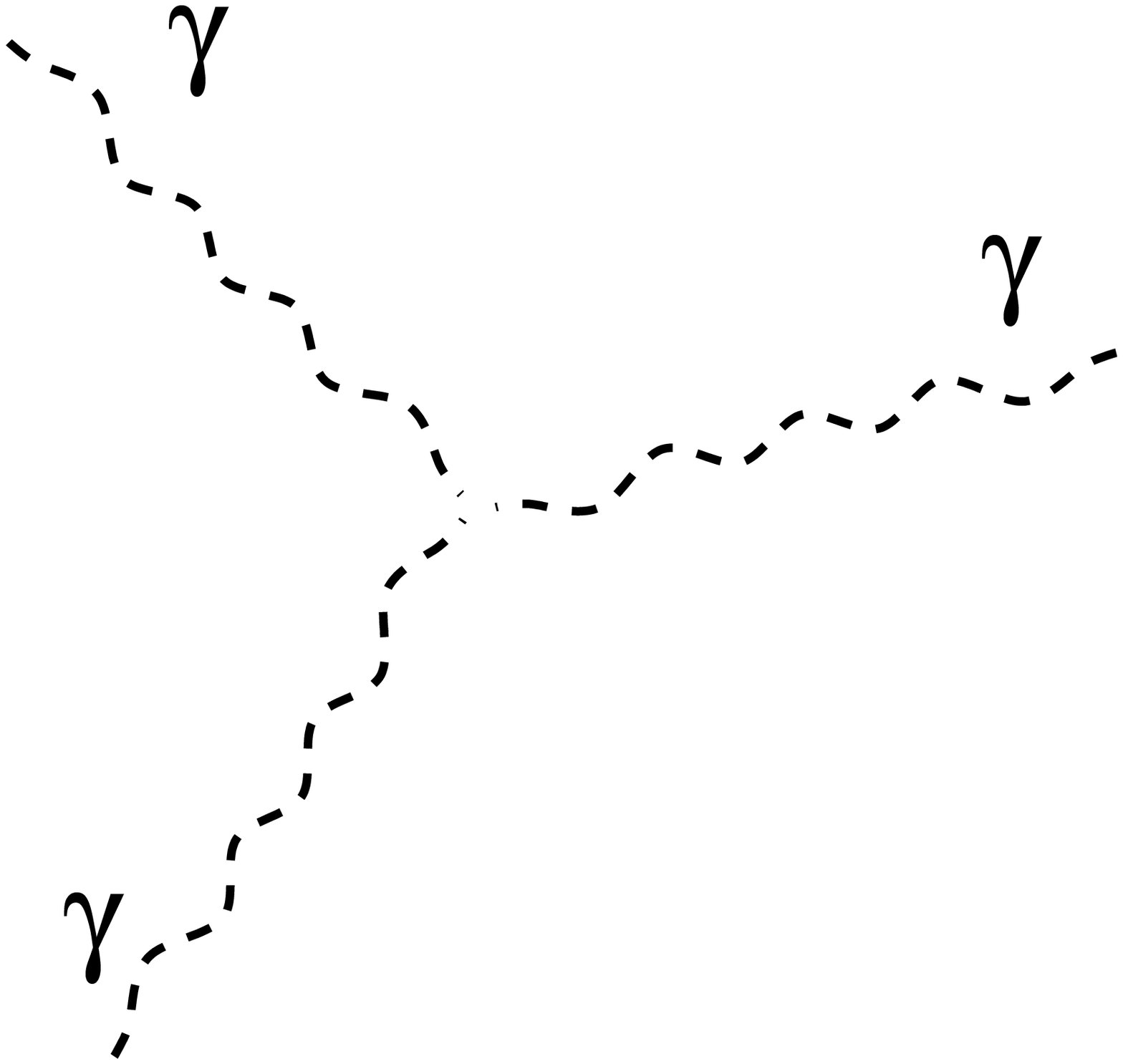}}} \\ \hline
Jets & Semi-leptonic & Leptonic & \multicolumn{2}{|c|}{3 photons}\\ \hline
\hline
\end{tabular}
\end{center}
\end{table}

\newpage

For its presumably final year of running LEP aims at collecting data at 
energies up to 207~GeV (optimistically up to 209~GeV). The 
foreseen sensitivity of the standard model Higgs boson
searches are expected
to be between 112 and 115~\Gcs. The 3$\sigma$ evidence sensitivity 
for a standard model
Higgs boson lies 300~MeV/c$^2$ below the 95\% CL exclusion and the 5$\sigma$
discovery is 2~\Gcs\ below this limit. A sizeable window is thus
still open at LEP, in a region where, if it exists, the 
standard model Higgs boson is most expected~\cite{sandra}. 

\section*{Acknowledgements}
It is a pleasure to thank the organisers and the secretariat of the 
XXXV$^{\rm th}$ Rencontres de Moriond for their 
wonderful hospitality. I would like to thank all my colleagues from the
four LEP collaborations and especially
Vanina Ruhlman Kleider, Sofia Andringa Dias, Andr\'e Tilquin, 
Jean-Baptiste de Vivie, Patrick Janot,  
Satoru Yamashita, Peter Igo-Kemenes and the LEP Higgs working group for their
help in preparing this talk. I also wish to express my gratitude to 
Ga\"{e}lle 
Boix, Jean-Ba and Patrick for their careful reading of these proceedings.

\end{document}